\documentclass[a4paper,12pt]{article}
\usepackage{chngcntr} % chngcntrパッケージを読み込む
\counterwithin{equation}{section} % セクションごとに数式番号をリセット

\usepackage[dvipdfmx]{graphicx}
\usepackage{authblk}
\usepackage{bm}
\usepackage{color}
\usepackage{array}
\usepackage{fancyhdr}
\usepackage{enumitem}
\usepackage{hyperref}
\usepackage{amsmath,amssymb}
  \usepackage{tabularx}
  \usepackage{cite}
  \usepackage{hhline}
   \newcolumntype{C}{>{\centering\arraybackslash}X}
   \newcolumntype{L}{>{\raggedright\arraybackslash}X}
   \newcolumntype{R}{>{\raggedleft\arraybackslash}X}
\setlistdepth{10}
\usepackage[font=small]{caption}

\renewcommand{\thesection}{\Roman{section}}

\renewcommand{\theequation}{\arabic{section}.\arabic{equation}}

\usepackage{titlesec}

\titleformat{\section}  % which section command to format
  {\fontsize{14}{16}\bfseries} % format for whole line
  {\Roman{section}.} % how to show number
  {0.5em} % space between number and text
  {} % formatting for just the text
  [] % formatting for after the text

\titleformat{\subsection}  % which section command to format
  {\fontsize{12}{14}\bfseries} % format for whole line
  {\Alph{subsection}.} % how to show number
  {0.5em} % space between number and text
  {} % formatting for just the text
  [] % formatting for after the text

  \titleformat{\subsubsection}  % which section command to format
  {\fontsize{10}{12}\bfseries} % format for whole line
  {\arabic{subsubsection}.} % how to show number
  {0.5em} % space between number and text
  {} % formatting for just the text
  [] % formatting for after the text

\usepackage[normalem]{ulem}

\newcommand{\ii}{\mathrm{i}}
\newcommand{\dd}{\mathrm{d}}

\newcommand{\del}{\partial}

\newcommand{\tb}{t_{\rm B}}

\definecolor{DarkBlue}{rgb}{0,0,0.7} 

\definecolor{DarkRed}{rgb}{0.65,0,0}

\definecolor{mycustomgreen}{RGB}{0, 120, 60}

\newcommand{\repnuma}{NU-QG-28}
\newcommand{\repnumb}{RUP-26-21}
\newcommand{\emaila}{\sf yoo.chulmoon.k6@f.mail.nagoya-u.ac.jp}

\fancypagestyle{titlepage}{%
  \fancyhf{}
  \fancyhead[R]{%
    \repnuma\\
    \repnumb\\
  }
   \fancyfoot[L]{%
    \rule{0.2\textwidth}{0.4pt} \\[-\baselineskip]%
    \vspace{4mm}
    \emaila}

}

\title{\bf \large Simulation of PBH formation in a matter-dominated universe}

\author[1,2]{\normalsize Chul-Moon~Yoo}
\author[1,3,4]{\normalsize Albert~Escriv\`a}
\author[5]{\normalsize Tomohiro~Harada}
\author[6,7,8,9]{\normalsize Kazunori~Kohri}
\affil[1]{\scriptsize \em Graduate School of Science, Nagoya University, Nagoya 464-8602, Japan}
\affil[2]{\scriptsize \em Kobayashi-Maskawa Institute for the Origin of Particles and the Universe (KMI),
Nagoya 464-8602, Japan}
\affil[3]{\scriptsize \em Asia Pacific Center for Theoretical Physics, Pohang 37673, Republic of Korea}
\affil[4]{\scriptsize \em Department of Physics, Pohang University of Science and Technology, Pohang 37673, Republic of Korea}
\affil[5]{\scriptsize \em Department of Physics, Rikkyo University, Toshima, Tokyo 171-8501, Japan}
\affil[6]{\scriptsize \em Division of Science, NAOJ, and SOKENDAI, 2-21-1 Osawa, Mitaka, Tokyo 181-8588, Japan}
\affil[7]{\scriptsize \em Department of Astronomy, The University of Tokyo, Bunkyo-ku, Hongo, Tokyo 113-0033, Japan}
\affil[8]{\scriptsize \em Theory Center, IPNS, KEK, 1-1 Oho, Tsukuba, Ibaraki 305-0801, Japan}
\affil[9]{\scriptsize \em Kavli IPMU (WPI), UTIAS, The University of Tokyo, Kashiwa, Chiba 277-8583, Japan}
\begin{document}
\baselineskip5.5mm
\date{}
%\if0
\maketitle
\thispagestyle{titlepage}

\begin{abstract}
\baselineskip5.5mm
\normalsize
We investigate primordial black hole (PBH) formation during an early matter-dominated era using fully nonlinear numerical relativity. 
The initial condition is set by a functional form of the curvature perturbation including 
ellipticity, which makes the configuration triaxial. 
Two kinds of matter descriptions are considered: dust fluid and collisionless particles. 
In the dust fluid description, numerical computation crashes associated with the appearance of a singularity at which the fluid density diverges, unless the singularity is hidden well inside the apparent horizon. 
We found that, for the dust fluid description, to observe the horizon formation before 
calculations crash, the initial amplitude must be larger than the previous analytic estimation by a factor of 2. 
On the other hand, with the particle system description, calculations do not crash, and 
we may observe black hole formation after subsequent evolution of the system. 
Then the threshold of black hole formation is significantly smaller than the previous analytic estimation by an order of magnitude.  

\end{abstract}

%\fi
%\baselineskip5.5mm

% \vspace{5cm}
% \thispagestyle{empty}
\pagebreak

\pagestyle{plain}
%%%%%%%%%%%%%%%%%%%%%%%%%%%%%%%%%%%%%%%%%%%%%%%%%%%%%%%%%%%%%%%%
\section{Introduction}
%%%%%%%%%%%%%%%%%%%%%%%%%%%%%%%%%%%%%%%%%%%%%%%%%%%%%%%%%%%%%%%%

Primordial black holes (PBHs)~\cite{1967SvA....10..602Z,Carr:1974nx,Hawking:1971ei} are black holes that may have formed in the early universe from the collapse of large density fluctuations. Unlike black holes produced by stellar collapse, PBHs can have a mass substantially smaller than the solar mass at the time of formation. 
The possible mass range of PBHs is wide, and many kinds of observational constraints 
have been discussed depending on the mass scale. 
Useful reviews in different aspects and a summary of observational constraints 
can be found in, e.g., Refs.~\cite{Carr:2009jm,Sasaki:2018dmp,Carr:2020gox,Green:2020jor,Carr:2020xqk,Escriva:2022duf,Yoo:2022mzl,Carr:2023tpt,Carr:2026hot}.

The possibility of PBH formation was first proposed more than fifty years ago, and the topic has gained renewed attention because PBHs could account for a fraction of dark matter~\cite{Carr:2016drx,Carr:2020gox}, seed supermassive black holes~\cite{Kawasaki:2012kn,Kohri:2014lza,Nakama:2016kfq}, black hole binaries observed by gravitational waves~\cite{Sasaki:2016jop,Bird:2016dcv} or possible microlensing objects~\cite{Niikura:2017zjd,Niikura:2019kqi,Sugiyama:2026kpv}~(see also Refs.~\cite{Mroz:2024wia,Mroz:2026nez}).

% generate observable signatures through gravitational waves and evaporation. In particular, scalar-induced gravitational waves from enhanced primordial curvature perturbations and collapse-induced gravitational waves from PBH formation offer complementary probes of the early universe. Observational constraints from cosmic microwave background anisotropies, microlensing surveys, gravitational wave detections, and gamma-ray backgrounds limit the allowed PBH abundance, but viable windows remain, especially in non-standard cosmological histories.

The formation of PBHs is highly sensitive to the equation of state of the dominant component in the universe. In a radiation-dominated era, pressure gradients oppose gravitational collapse, requiring large perturbations to form PBHs. 
According to the peak statistics \cite{Bardeen:1985tr}, a large-amplitude peak tends to be spherically symmetric. 
This suggests that, consistently with the numerical explorations performed so far, spherical symmetry may provide a good approximation in some settings (see Refs.~\cite{Yoo:2020lmg,Yoo:2024lhp,Escriva:2024aeo,Escriva:2024lmm} for supporting results from numerical simulations).
% the spherically symmetric assumption is considered to be valid 
% \AEE{I personally think is strong, what about 
% "This suggests that spherical symmetry may provide a good approximation in some settings, consistently with the numerical explorations performed so far"}  
In contrast, during a matter-dominated epoch, pressure is negligible and overdense regions can collapse more easily in general. 
Such early matter-dominated phases can arise from the oscillation of inflaton or other scalar fields after inflation before reheating completes.
As is discussed in several previous works \cite{Harada:2016mhb,Harada:2017fjm,Saito:2024hlj,Ye:2025wif} (see Refs.~\cite{1980PhLB...97..383K,1982SvA....26....9P,Kokubu:2018fxy} and Ref.~\cite{Harada:2022xjp} for effects of inhomogeneity and velocity dispersion, respectively), non-spherically symmetric dynamics would be essential for the PBH formation criterion. 

One interesting aspect associated with the non-spherical dynamics is the generation of gravitational waves. 
Gravitational wave generation from gravitational collapse and structure formation in an early matter-dominated era has been discussed in Refs.~\cite{2010JCAP...04..021J,PhysRevD.101.063519,Eggemeier:2022gyo,Flores:2022uzt,Eggemeier:2023nyu,Fernandez:2023ddy,Zeng:2025law,Dalianis:2020gup,Dalianis:2024kjr,Escriva:2026blk}, while perturbative calculation of induced gravitational waves can be seen in 
Refs.~\cite{Assadullahi:2009nf,Kohri:2018awv,Inomata:2019zqy,PhysRevD.100.043532,Pearce:2023kxp,Kumar:2024hsi}. 
In general, to obtain a sizable amplitude of the background gravitational wave spectrum induced by scalar modes, an enhancement mechanism of primordial scalar perturbations is needed. Since the enhancement of the primordial scalar perturbation inevitably enhances PBH formation probability, PBH abundance and induced gravitational waves are often discussed together. 
However, PBH formation during matter-dominated era is not as well understood as it is during the radiation-dominated era, although there are some analytic works with not necessarily valid perturbative approximations and hypotheses~\cite{Harada:2016mhb,Harada:2017fjm,Saito:2024hlj,Ye:2025wif,1980PhLB...97..383K,1982SvA....26....9P,Kokubu:2018fxy,Harada:2022xjp}. 

The main difficulty in the investigation into the PBH formation in a matter-dominated epoch is 
the implementation of nonspherical symmetry in the numerical simulation. 
The numerical relativity techniques have been developed over the past 30 years 
since the formulation of the Baumgarte-Shapiro-Shibata-Nakamura (BSSN)  formalism and have become a well-established tool to investigate 
relativistic non-linear gravitational dynamics. 
However, numerical relativity beyond spherical symmetry has been mainly developed in asymptotically flat settings, and 
the use of 3+1 dimensional numerical relativity in PBH formation is still under development (see \cite{Yoo:2012jz,Yoo:2013yea,Yoo:2014boa,Yoo:2018pda} for related early works on the system including a black hole in an expanding box). 
Nevertheless, 3+1 dimensional numerical investigation into black hole systems in an expanding background has been developing rapidly in recent times~\cite{deJong:2021bbo,deJong:2023gsx,Yoo:2020lmg,Yoo:2024lhp,Escriva:2024lmm,Escriva:2024aeo,Kitajima:2025shn,Baumgarte:2026igz,Escriva:2026kov}. 
The recent developments enable us to perform the simulation of PBH formation for the systems filled with fluid and scalar fields. 
Even for the pressureless dust fluid, we can perform 3+1 dimensional simulation in an expanding background~(see, e.g., Ref.~\cite{Munoz:2023rwh}). 
However, for a dust fluid, we suffer from the crash of the numerical computation associated with the shell-crossing singularity, which 
generically happens due to the intersection of fluid elements. 
Therefore, we cannot probe the spacetime after the moment of the shell crossing with the dust fluid description. 

In this work, we focus on PBH formation in a matter-dominated universe using fully nonlinear numerical relativity. 
Focusing on the effects of inhomogeneity~\cite{1980PhLB...97..383K,1982SvA....26....9P,Kokubu:2018fxy} and ellipticity~\cite{Harada:2016mhb}, 
we compare our numerical results with previous analytic investigations. 
Except for the last part, the dust fluid description is adopted, and we discuss PBH formation 
relying on the results before computation crashes due to the singular behavior of the density distribution. 
This singular behavior might best be interpreted as arising from the breakdown of the fluid approximation. 
Therefore, in the last part, as a possible resolution, we try to simulate the same system adopting the description of the collisionless particle system. 
That is, we show the results of the attempt to perform a particle-in-cell (PIC) simulation in full numerical relativity for PBH formation.

This paper is organized as follows. In Section~\ref{sec:bssn}, we summarize the BSSN formulation used in our simulations and introduce the matter variables. Section~\ref{sec:gaugecond} discusses the gauge conditions and the scale-up non-Cartesian coordinate system adopted for the numerical integration. Section~\ref{sec:initial} describes the construction of the initial data for the dust fluid. 
%  and collisionless particle systems. 
In Section~\ref{sec:sph}, we review spherically symmetric cases, including the LTB spacetime, the correspondence with long-wavelength solutions, the nakedness of the singularity, and spherical dust fluid simulations. Section~\ref{sec:asph} presents the aspherical collapse setup, the Zel'dovich approximation, and the comparison between analytic estimates and our numerical results. Section~\ref{sec:cps} introduces the collisionless particle system, the stress-energy tensor and particle settings, and reports the particle-based simulation results. Finally, Section~\ref{sec:conclusion} summarizes our conclusions and outlook. We use geometric units where $G = c = 1$ throughout the paper.
%%%%%%%%%%%%%%%%%%%%%%%%%%%%%%%%%%%%%%%%%%%%%%%%%%%%%%%%%%%%%%%%
\section{Basic setups in the form of 3+1 decomposition}
\label{sec:basic_eqs}
%%%%%%%%%%%%%%%%%%%%%%%%%%%%%%%%%%%%%%%%%%%%%%%%%%%%%%%%%%%%%%%%

\subsection{Basic equations in BSSN formalism}
\label{sec:bssn}

To fix the notation, let us start by introducing geometrical variables. 
We consider the following form of line elements:
\begin{equation}
\dd s^2=-\alpha^2 \dd t^2+\gamma_{ij} 
	\left(\dd x^i+\beta^i \dd t\right)
	\left(\dd x^j+\beta^j \dd t\right), 
\end{equation}
where $\gamma_{ij}$, $\alpha$ and $\beta^i$ are 
the spatial metric, lapse function, and shift vector, respectively. 
For numerical integration, we decompose the spatial metric as 
\begin{equation}
\gamma_{ij}=a^2\psi^4 \tilde \gamma_{ij}~~{\rm with}~~\det \tilde \gamma=\det f,  
\end{equation}
where $\det f$ is the determinant of the 3-dim flat metric $f_{ij}$. 
The unit normal vector field $n^\mu$ to the spatial hypersurface 
is given by $n_\mu=-\alpha\left(\dd t\right)_\mu$. 
Then, the projection tensor is given by 
\begin{equation}
\gamma_\mu^{~\nu}=n_\mu n^\nu+g_\mu^{~\nu}. 
\end{equation}
The extrinsic curvature $K_{ij}$ is defined by 
\begin{equation}
K_{ij}=-\gamma_i^{~\mu}\gamma_j^{~\nu} \nabla_\mu n_\nu. 
\end{equation}
We adopt the following decomposition of the extrinsic curvature:
\begin{equation}
K_{ij}=a^2\psi^4 \tilde A_{ij}+\frac{1}{3}K\gamma_{ij}. 
\end{equation}
In the numerical integration, 
$\alpha$, $\beta^i$, $\tilde \gamma_{ij}$, $\psi$, $K$ and $\tilde A_{ij}$
are regarded as independent variables. 

As for matter variables, for convenience, we introduce 
the following independent variables:
\begin{eqnarray}
E&=&n^\mu n^\nu T_{\mu\nu}, 
\label{eq:rhon}
\\
J^i&=&-\gamma^i_\mu n_\nu T^{\mu\nu}, 
\label{eq:ji}
\\
S_{ij}&=&\gamma_i^{~\mu}\gamma_j^{~\nu}T_{\mu\nu}, 
\label{eq:sij}
\end{eqnarray}
where $T_{\mu\nu}$ is the energy momentum tensor. 

Following the conventional method, called the BSSN formalism\cite{Shibata:1995we,Baumgarte:1998te}, 
we obtain evolution equations for $\tilde \gamma_{ij}$, $\psi$, $K$ and $\tilde A_{ij}$ 
as follows:
\begin{eqnarray}
  (\partial_{t}-{\cal L}_{\beta})\tilde{\gamma}_{ij}&=&-2\alpha
   \tilde{A}_{ij}-\frac{2}{3}\tilde{\gamma}_{ij}{\cal D}_{k}\beta^{k}, 
 \label{eq:SS99_2.11}\\
  (\partial_{t}-{\cal L}_{\beta})\tilde{A}_{ij}&=&\frac{1}{a^{2}\psi^{4}}
 \left[\alpha\left({\cal R}_{ij}-\frac{\gamma_{ij}}{3}{\cal R}\right)
 -\left(D_{i}D_{j}\alpha-\frac{\gamma_{ij}}{3}D_{k}D^{k}\alpha\right)\right]
 \nonumber \\
 && +\alpha(K\tilde{A}_{ij}-2\tilde{A}_{ik}\tilde{A}_{j}^{k})
 -\frac{2}{3}({\cal D}_{k}\beta^{k})\tilde{A}_{ij}
 -\frac{8\pi\alpha}{a^{2}\psi^{4}}\left(S_{ij}-\frac{\gamma_{ij}}{3}S_{k}^{k}\right), 
 \label{eq:SS99_2.12}
 \\
 (\partial_{t}-{\cal L}_{\beta})\psi &=&
  -\frac{1}{2}H\psi+\frac{\psi}{6}(-\alpha K+{\cal D}_{k}\beta^{k}), \label{eq:SS99_2.13} \\
 (\partial_{t}-{\cal L}_{\beta})K &=&
  \alpha\left(\tilde{A}_{ij}\tilde{A}^{ij}+\frac{1}{3}K^{2}\right)-D_{k}D^{k}\alpha
  +4\pi\alpha (E+S_{k}^{k}), 
 \label{eq:SS99_2.14}
 \end{eqnarray}
% \begin{eqnarray}
% \left(\del_t-\beta^i \del_i\right)\psi
% &=&\frac{1}{6}\psi\left(\del_i\beta^i-\alpha K\right),
% \\
% \left(\del_t-\beta^k \del_k\right)\tilde \gamma_{ij}
% &=&-2\alpha \tilde A_{ij}+\tilde \gamma_{ik} \del_j\beta^k
% +\tilde \gamma_{jk}\del_i\beta^k
% -\frac{2}{3}\del_k \beta^k \tilde \gamma_{ij},
% \\
% \left(\del_t -\beta^i\del_i\right)K
% &=&\alpha\left[\tilde A_{ij} \tilde A^{ij}+\frac{1}{3}K^2
% +4\pi\left(E+S\right)\right]-D^iD_i\alpha, 
% \label{eq:dtk1}
% \\
% \left(\del_t -\beta^k\del_k\right)\tilde A_{ij}
% &=&\ee^{-4\psi}
% \left(\alpha\mathcal R_{ij}-D_iD_j\alpha-8\pi\alpha S_{ij}\right)^{\rm T}
% \cr&&
% +\alpha\left(K\tilde A_{ij}-2\tilde A_{ik}\tilde A^k_{~j}\right)
% +\tilde A_{kj}\del_i\beta^k+\tilde A_{ik}\del_j \beta^k
% -\frac{2}{3}\tilde A_{ij}\del_k\beta^k, 
% \end{eqnarray}
  where $D_i$ and ${\mathcal D}_i$ are the covariant derivatives for $\gamma_{ij}$ and the 3-dim flat metric $f_{ij}$, respectively, 
  and $\mathcal R_{ij}$ is the Ricci tensor for $\gamma_{ij}$. 
  $H=(\partial_t a)/a$ and 
  $S_{ij}$ is defined by $\gamma_{i\mu}\gamma_{j\nu}T^{\mu\nu}$. 
% The Ricci tensor $\mathcal R$ is calculated by the following expression:
% \begin{eqnarray}
% \mathcal R_{ij}&=&\tilde {\mathcal R}_{ij}
% -2\tilde D_i \tilde D_j \psi+4\tilde D_i\psi \tilde D_j \psi 
% -2\tilde \gamma_{ij}\tilde D^k\tilde D_k \psi 
% -4\tilde \gamma_{ij}\tilde D_k \psi \tilde D^k\psi, \\
% \tilde {\mathcal R}_{ij}&=&
% -\frac{1}{2}\del_l\tilde \gamma_{ij}\tilde \Gamma^l_0
% +\frac{1}{2}\del_j\tilde \Gamma^l\tilde \gamma_{li}
% +\frac{1}{2}\del_i\tilde \Gamma^l\tilde \gamma_{lj}
% -\frac{1}{2}\tilde\gamma^{kl}\del_l\del_k\tilde \gamma_{ij}
% \cr&&\hspace{5cm}
% +\tilde\Gamma^{kl}_{~~i}\tilde\Gamma_{jkl}+\tilde\Gamma^{kl}_{~~j}\tilde\Gamma_{ikl}
% +\tilde\Gamma^{kl}_{~~i}\tilde\Gamma_{klj}, 
% \end{eqnarray}
% where $\tilde \Gamma^i_{~jk}=\tilde\gamma_{jl}\Gamma^{il}_{~~k}=\tilde\gamma^{il}\Gamma_{ljk}$ is the Christoffel symbol with respect to $\tilde \gamma_{ij}$, and 
% $\tilde \Gamma^i_0=\tilde \Gamma^i=-\del_j\tilde \gamma^{ij}$. 
% For stable numerical integration, we calculate $\tilde \Gamma^i$ by the following 
% evolution equation:
% \begin{eqnarray}
% \left(\del_t-\beta^j\del_j\right)\tilde \Gamma^i
% &=&\tilde \gamma^{jk}\del_j\del_k\beta^i
% +\frac{1}{3}\tilde \gamma^{ij}\del_j\del_k\beta^k
% -\tilde \Gamma^j\del_j\beta^i
% +\frac{2}{3}\tilde\Gamma^i\del_j\beta^j
% -2\tilde A^{ij}\del_j\alpha
% \cr&&
% +2\alpha\left(\tilde \Gamma^i_{jk}\tilde A^{jk}
% +6\tilde A^{ij}\del_j\psi
% -\frac{2}{3}\tilde \gamma^{ji}\del_jK-8\ee^{4\psi}\pi J^i\right). 
% \end{eqnarray}
Then the Hamiltonian and momentum constraint equations can be written as follows:
\begin{eqnarray}
&&  \tilde \triangle \psi=\frac{1}{8}\tilde {\mathcal R}^k_k\psi-2\pi \psi^5a^2E-\frac{1}{8}\psi^5 a^2\left(\tilde A_{ij}\tilde A^{ij}-\frac{2}{3}K^2\right), 
\label{eq:hamcon}\\
&&\tilde {\mathcal D}^j(\psi^6\tilde A_{ij})-\frac{2}{3}\psi^6\tilde {\mathcal D}_i K=8\pi J_i\psi^6,  
\label{eq:momcon}
\end{eqnarray}
where $\tilde \triangle$, $\tilde {\mathcal D}_i$ and $\tilde {\mathcal R}_{ij}$ are the Laplacian, covariant derivative and the Ricci tensor for $\tilde \gamma_{ij}$, respectively. 
We also consider dynamical equations for $\alpha$ and $\beta^i$ to fix the gauge modes, which will be discussed later.

%%%%%%%%%%%%%%%%%%%%%%%%%%%%%%%%%%%%%%%%%%%%%%%%%%%%%%%%%%%%%%%%
% \section{Gauge conditions}
%%%%%%%%%%%%%%%%%%%%%%%%%%%%%%%%%%%%%%%%%%%%%%%%%%%%%%%%%%%%%%%%

\subsection{Gauge conditions}
\label{sec:gaugecond}

In this work, we use a novel time slicing condition specialized for simulation of PBH formation (see Ref.~\cite{Ning:2026jkk} for a similar approach). 
To distinguish different definitions of the lapse function, we denote the lapse function associated with the standard cosmological time $t$ as $\alpha_{\rm c}$. 
In PBH formation, two characteristic time scales exist. 
One is given by $1/H$, 
% \THc{$H$ is undefined.}, 
which is time-dependent, and the other is given by the timescale at the horizon entry 
$1/H_k$, which is characterized by the scale of the inhomogeneity $k$ and is time-independent. 

First, for simplicity, let us consider the homogeneous background. 
For the background, 
to efficiently resolve the cosmological dynamics, the e-folding number $N$ 
defined by 
\begin{equation}
  N:=\ln \frac{a}{a_{\rm ini}}
\end{equation}
is useful because an interval $\Delta N$of $N$ is the cosmological time interval normalized by $H$, namely, 
$\Delta N=H\Delta t$. 
Therefore, it might be convenient to use $N$ instead of the cosmological time $t$. 
To keep the dimension of the time coordinate, for convenience, let us consider the time coordinate $T:=N/H_k$. 
In this case, the lapse function $\alpha_{\rm c}$ should be transformed to $\alpha_T$ as follows:
\begin{equation}
\alpha_{\rm c} \dd t=\alpha_{\rm c}\frac{H_k}{H}\dd T=\alpha_T\dd T.  
\end{equation}
From this equation, we obtain 
% \AEE{eq.2.16 explicitly defines the transformed lapse $\alpha_T$ but then eq.2.21 then calls the transformed initial quantity simply $\alpha$ while eq.3.8 gives the cosmological-time lapse under the same symbol...i am confused with that}
\begin{equation}
  \frac{\dd \alpha_{\rm c}}{\dd t}=\left(\frac{H}{H_k}\right)^2\frac{\dd\alpha_T}{\dd T}+\frac{H}{H_k^2}\frac{\dd H}{\dd T}\alpha_T=\left(\frac{H}{H_k}\right)^2\left(\frac{\dd\alpha_T}{\dd T}-\frac{3}{2}H_k\alpha_T\right). 
\end{equation}
Therefore, for $\dd\alpha_{\rm c}/\dd t=0$, $\alpha_T$ exponentially grows as $\alpha_T\sim\exp\left(3H_k T/2\right)$. 
% \THc{$\alpha_N$? $\alpha_T$?}

Once we introduce an inhomogeneity, the e-folding number is not appropriate for resolving gravitational collapse after the horizon entry 
since the dynamical timescale is given by $1/H_k$. 
In this regime, a dynamical slicing condition with a singularity avoidance property is appropriate. 
For this reason, in simulations of PBH formation, for instance, the following modified version of the 1+log slicing condition is often used:
\begin{equation}
  \left(\del_t-\beta^i\del_i\right)\alpha_{\rm c}=-2\alpha_{\rm c}\left(K+3H\right), 
\end{equation}
where $H$ is the Hubble expansion rate for the background FLRW spacetime, 
and is equivalent to the value of $-K/3$ evaluated at the boundary of the numerical box. 
In this gauge condition, at the boundary, since the value of $\alpha_{\rm c}$ is fixed, 
the time coordinate is equivalent to the cosmological time of the background spacetime. 

An efficient slicing condition that can save computational time may be 
given by considering a gauge condition that initially resembles the e-folding number, and approaches to the cosmological time somewhat before the horizon entry. 
The background scale factor at the horizon entry of the scale $1/k$ is given by 
\begin{equation}
  a_k=\left(\frac{k}{a_\ii H_\ii}\right)^{-2/(1+3w)}a_i. 
\end{equation}
Therefore, we may consider this value as a reference value of $\alpha$ 
making the physical scales of the spatial and temporal coordinates comparable around the horizon entry time. 
Based on the above considerations, we adopt the following time slicing condition:
\begin{equation}
  \left(\del_t-\beta^i\del_i\right)\alpha=-2\alpha\left(K+3H\right)+\frac{3}{2}\alpha H_k\exp\left(-C^2\alpha_{\rm b}^2/a_k^2\right), 
\end{equation}
where $\alpha_{\rm b}$ is the value of the lapse function at the corner of the numerical box $x=y=z=L$, 
and we fixed the constant $C$ as $5$ after some numerical trials. 
The initial profile of $\alpha$ is set as 
\begin{equation}
\alpha=\frac{H_k}{H_\ii}-\frac{H_k}{3H_\ii}q\left(\frac{1}{a_\ii H_\ii}\right)^2. 
\end{equation}

As for the shift vector, we use a slightly modified version of 
the Hyperbolic Gamma driver given by 
\begin{eqnarray}
\left(\del_t-\beta^i\del_i\right) \beta^i&=&\frac{3}{4}B^i,\\
\left(\del_t-\beta^i\del_i\right) B^i&=&\del_t \tilde \Gamma^i -3H B^i,  
\end{eqnarray}
where $\tilde \Gamma^i=-\mathcal D_j \tilde \gamma^{ij}$.

\subsection{Scale-up non-Cartesian coordinates}
\label{sec:nonCart}

In this work, we consider the region $0\leq X^i \leq L$ with $X^i$ ($i=1,2,3$) being Cartesian coordinates. 
Then we employ the non-Cartesian coordinate system $(x,y,z)$, 
first implemented in Ref.~\cite{Yoo:2018pda}. 
The non-Cartesian coordinates $x^i$ are related to the Cartesian coordinates $X^i$ as follows:
\begin{equation}
X^i=x^i-\frac{\eta}{1+\eta}\frac{L}{\pi}\sin\left(\frac{\pi}{L}x^i\right).
\end{equation}
This functional form is compatible with the boundary conditions adopted in this work and satisfies $x^i=0$ at $X^i=0$ and $x^i=L$ at $X^i=L$.

At the origin, the infinitesimal interval in the Cartesian coordinate $\Delta X$ is covered by the non-Cartesian coordinate interval $\Delta x=(1+\eta)\Delta X$.
Therefore, the central part is enlarged in this non-Cartesian coordinate system. 
The value of $\eta$ is set to $10$.

%%%%%%%%%%%%%%%%%%%%%%%%%%%%%%%%%%%%%%%%%%%%%%%%%%%%%%%%%%%%%%%%
\section{Initial condition}
\label{sec:initial}
%%%%%%%%%%%%%%%%%%%%%%%%%%%%%%%%%%%%%%%%%%%%%%%%%%%%%%%%%%%%%%%%
% \subsection{Initial matter distribution}
We set the initial distribution based on the corresponding dust fluid description. 
Thus, we first consider the dust fluid stress-energy tensor given by 
\begin{equation}
 T^{\mu\nu}=\rho  u^\mu u^\nu. 
 \label{eq:dusttmn}
\end{equation}
% where we have identified the fluid four-velocity to the particle four-velocity. 
Then, $E$ and $J^\mu$ are given by 
\begin{eqnarray}
  E&=&\Gamma^2\rho, \label{eq:E}\\
  J^\mu&=&EV^\mu. \label{eq:J}
\end{eqnarray}

Since we are interested in PBH formation, long-wavelength growing mode solutions are appropriate as initial conditions. 
Under the fluid description, analytic expressions for long-wavelength solutions have been derived in Refs.~\cite{Shibata:1999zs,Harada:2015yda} up through 
next-to-leading order of the gradient expansion for a given functional form of the curvature perturbation $\zeta$. 
Specifically, using the function 
% $\Psi(\bm x):=\ln\zeta /2$ 
$\Psi(\bm x):=\exp\left(\zeta /2\right)$, we express the initial conditions for the geometrical variables as follows:
\begin{eqnarray}
  K&=&-3H,\\
\psi&=&
  % \xi&=&
  \Psi\left[1 
  -\frac{1}{6}q\left(\frac{1}{aH}\right)^2\right],\\
\tilde \gamma_{ij}&=&
  % h_{ij}&=&
  f_{ij}-\frac{4}{5}p_{ij}\left(\frac{1}{aH}\right)^{2}, 
 \label{eq:h_ij_solution}\\
 \tilde{A}_{ij}&=&
 \frac{2}{5}p_{ij}H
 \left(\frac{1}{aH}\right)^{2}, 
 \label{eq:A_tilde_ij_solution}\\
 \alpha_c&=&  
% \chi&=&
1-\frac{1}{3}q\left(\frac{1}{aH}\right)^2,\\ 
\beta^i&=&0,  
 \end{eqnarray}
where  
% \AEE{for consistency with the other eqaution, i think the definition should be $q(\bm x)=-\frac{4}{3}\frac{\triangle \Psi}{\Psi^5}$, am i wrong?}\CY{right, typo}
\begin{eqnarray}
  q(\bm x)&=&-\frac{4}{3}\frac{\triangle \Psi}{\Psi^5}, \\
  p_{ij}(\bm x)&=&\frac{1}{\Psi^{4}}
 \left[
 -\frac{2}{\Psi}\left({\cal D}_{i}{\cal D}_{j}\Psi-\frac{1}{3}f_{ij}\triangle\Psi\right)+\frac{6}{\Psi^{2}}\left({\cal D}_{i}\Psi{\cal D}_{j}\Psi-\frac{1}{3}f_{ij}{\cal D}^{k}\Psi{\cal D}_{k}\Psi\right)
 \right]
 \end{eqnarray}
 with $\triangle:=f^{ij}\mathcal D_i\mathcal D_j$. 
 % \AEE{equation 3.6 for the factor 4/5 has a correct sign?looks needed a magnus (-) following consistency with eqs(2.9) and (3.7)}\CY{corrected} 
The initial hypersurface is a constant-mean-curvature slice, since the value of $K$ is constant. 
Once the geometrical variables are given, through the constraint equations \eqref{eq:hamcon} and \eqref{eq:momcon}, we can calculate $E$ and $J^\mu$. 
Then, we obtain $\rho$ and $V^\mu$ from Eqs.~\eqref{eq:E} and \eqref{eq:J}. 

In the numerical simulations presented in this paper, we set $H_i=5k=50/L$ and $a_{\rm i}=1$. 

%%%%%%%%%%%%%%%%%%%%%%%%%%%%%%%%%%%%%%%%%%%%%%%%%%%%%%%%%%%%%%%%
\section{Spherically symmetric cases}
\label{sec:sph}
%%%%%%%%%%%%%%%%%%%%%%%%%%%%%%%%%%%%%%%%%%%%%%%%%%%%%%%%%%%%%%%%

In this section, for a test calculation, 
we compare our simulation with a dust fluid 
to the corresponding LTB solution. 

\subsection{LTB spacetime}

Line elements in the LTB spacetime can be expressed as 
\begin{equation}
\dd s^2=-\dd t^2+\frac{(\del_r R)^2}{1-k(r)r^2}\dd r^2+R(t,r)^2\dd \Omega^2, 
\label{eq:LTBmetric}
\end{equation}
where $\tilde k$ is an arbitrary function of the radial coordinate $r$ and 
the area radius $R$ is a function of $t$ and $r$. 
Hereafter, we omit the argument unless it is confusing. 
The energy-momentum tensor is given by the dust fluid form as Eq.~\eqref{eq:dusttmn} with $u^\mu=(\del_t)^\mu$. 
The Einstein equations lead to 
\begin{equation}
(\del_t R)^2=-\tilde kr^2+\frac{m(r)r^3}{3R}, 
\label{eq:ltb}
\end{equation}
where $m$ is an arbitrary function of $r$ and 
related to the energy density $\rho$ as 
% \begin{equation}
% \rho=\frac{\del_rmr^3+3mr^2}{24\pi\del_r R R^2}. 
% \end{equation}
\begin{equation}
\rho=\frac{(\del_rm)r^3+3mr^2}{24\pi R^2\del_r R }. 
\end{equation}
The Misner-Sharp mass\cite{Misner:1964je} is given by 
\begin{equation}
M=\frac{1}{6}mr^3. 
\end{equation} 

We express the solution of Eq.~\eqref{eq:ltb} 
following the convention proposed in Ref.~\cite{Tanimoto:2007dq}
as follows: 
\begin{equation}
R=rm^{1/3}(t-\tb(r))^{2/3}S(x), 
\label{eq:ltbR}
\end{equation}
where $\tb$ is an arbitrary function of $r$, 
$x=\tilde k\left(\frac{t-\tb}{m}\right)^{2/3}$ and 
the function $S$ is defined by 
\begin{equation}
S(x)=
\left\{\begin{array}{lll}
\displaystyle
\frac{\cosh\sqrt{-\eta}-1}{6^{1/3}(\sinh\sqrt{-\eta}
-\sqrt{-\eta})^{2/3}}
\,;\qquad
&\displaystyle
x=\frac{-(\sinh\sqrt{-\eta}-\sqrt{-\eta})^{2/3}}{6^{2/3}}
\quad&\mbox{for}~~x<0\,,
\\
\displaystyle
\frac{1-\cos\sqrt{\eta}}{6^{1/3}(\sqrt{\eta}
-\sin\sqrt{\eta})^{2/3}}
\,;&\displaystyle
x=\frac{(\sqrt{\eta}-\sin\sqrt{\eta})^{2/3}}{6^{2/3}}
\quad&\mbox{for}~~x>0\,,
\end{array}\right.
\label{eq:defS}
\end{equation}
and $S(0)=({3}/{4})^{1/3}$. 
The function $S(x)$ is analytic for $x<(\pi/3)^{2/3}$. 

\subsection{Correspondence with long-wavelength solutions}

There are three arbitrary functions in the expression of the LTB solution. 
In this paper, we focus on the case $\tb(r)=0$ since 
inhomogeneity induced by the nontrivial functional form of $\tb(r)$ 
corresponds to decaying modes. 
Then, there are two functional degrees of freedom $\tilde k(r)$ and $m(r)$. 
These degrees of freedom correspond to growing modes and the gauge degree of freedom 
associated with the choice of the radial coordinate. 
For PBH formation, the relevant initial inhomogeneity is described by the solution in the long-wavelength limit, that is, 
the lower orders of the gradient expansion.

In the general framework of the 3+1 formulation, on the other hand, at the leading order of the gradient expansion, focusing on growing modes, 
one can describe the spacetime metric as \cite{Shibata:1999zs,Lyth:2004gb,Harada:2015yda}
\begin{equation}
\dd s^2=-dt^2+a^2\Psi^4\left(\dd r^2+r^2\dd\Omega^2\right), 
\label{eq:isomet}
\end{equation}
where 
$\Psi$ is given by a function of the radial coordinate $r$ with spherical symmetry. This leading-order form of the metric commonly applies to the time-slicing conditions of the constant-mean-curvature slice, the comoving slice, and the uniform density slice for adiabatic systems.
In this form, the radial coordinate is often called the isotropic coordinate, in which the spatial metric is described in a conformally flat form. 

Since $\Psi$ is time-independent at the leading order of the gradient expansion, which is valid at a sufficiently early time, the LTB metric without decaying modes should also be described by the form of Eq.~\eqref{eq:isomet} at a sufficiently early time. %$t=t_{\rm i}$. 
Then, 
let us compare the metric in Eqs.~\eqref{eq:isomet} and ~\eqref{eq:LTBmetric} 
%at 
% a sufficiently early time 
%$t=t_{\rm i}$ 
in the limit of $t\to 0$. 
More precisely, we can require the following equations:
%Comparing the spatial metrics, we obtain
%\dred{
\begin{eqnarray}
%a_{\rm i}\Psi^2&=&\frac{R|_{t=t_{\rm i}}}{ r}=\left(\frac{3m(r)}{4}\right)^{1/3}t_{\rm i}^{2/3}, 
r\Psi^{2}(r)&=&\lim_{t\to 0}\frac{R(t,r)}{a(t)}
\label{eq:Psi2}
\\
%a_{\rm i}^2\Psi^4&=&\frac{\left(\del_r R|_{t=t_{\rm i}}\right)^2}{1-\tilde kr^2},  
\Psi^4 (r) &=& \lim_{t\to 0}\frac{1}{1-\tilde kr^2}\left[\del_r \left(\frac{R (t,r)}{a(t)}\right)\right]^2.
\label{eq:Psi4}
\end{eqnarray}
%}
% \begin{eqnarray}
% \Psi^2&=&\frac{R}{a r}=\left(\frac{3m(r)}{4}\right)^{1/3}t_{\rm i}^{2/3}, 
% \label{eq:Psi2}
% \\
% \Psi^4&=&\frac{\del_r R^2}{1-\tilde kr^2},  
% \label{eq:Psi4}
% \end{eqnarray}
%where \dred{

We choose the background FLRW scale factor $a=a_{\rm i}(t/t_{\rm i})^{2/3}$ with 
$a_{\rm i}=a(t_{\rm i})$, where the time $t=t_{\rm i}$ is identified with the initial time considered in Sec.~\ref{sec:initial}. Thus we set $a_{\rm i}=1$, consistent with the parameter choice stated at the end of Sec.~\ref{sec:initial}. 
Noting $\lim_{x\to 0} S(x)=S(0)=(3/4)^{1/3}$,   
% and $a=\left(t/t_\ii\right)^{2/3}$.  
from the first equation \eqref{eq:Psi2},
we find 
\begin{equation}
  m(r)=\frac{4\Psi^6}{3 t_\ii^2}. 
  \label{eq:m}
\end{equation}
Taking the derivative of the first equation \eqref{eq:Psi2} with respect to $r$, we obtain 
\begin{equation}
%\del_r R|_{t=t_{\rm i}}=\Psi^2+2r\Psi\del_r\Psi. 
\Psi^2+2r\Psi\del_r\Psi=\del_r
\left(\lim_{t\to 0}\frac{R(t,r)}{a(t)}
\right).
\end{equation}
% \begin{equation}
% \del_r R=\Psi^2+2r\Psi\del_r\Psi. 
% \end{equation}
Substituting this equation into \eqref{eq:Psi4} and swapping the order of the $r$-derivative and the $t\to 0$ limit that is guaranteed for long-wavelength solutions, we obtain 
\begin{equation}
  \tilde k(r)=\frac{1}{r^2}\left[1-\left(1+2r\del_r\ln\Psi\right)^2\right]. 
  \label{eq:k}
\end{equation}
Therefore, the functional forms of $\tilde k(r)$ and $m(r)$ are uniquely given in terms of $\Psi(r)$ in the present setting.

\subsection{Nakedness of the singularity and PBH formation}
Hereafter, we consider the following functional form of $\Psi(r)$:
\begin{equation}
\ln \Psi=\frac{\mu}{2}\exp\left[-\frac{1}{6}k^2r^2\right]. 
\label{eq:Psi}
\end{equation}
The radius of the maximum compaction function $r_{\rm m}$ can be given by 
\begin{equation}
  \left. \left(\del_r +r\del_r^2\right)\ln \Psi\right|_{r=r_{\rm m}}=0\Leftrightarrow k^2r_{\rm m}^2=6.  
\end{equation}
Therefore, a typical scale is given by $ar=\sqrt{6}a/k$. 
Estimating the horizon entry by using this typical scale, we find 
\begin{equation}
\left.\frac{k^2}{a^2H^2}\right|_{\rm ent}=6. 
\end{equation}
Then, since $a^3H^3=2H_{\rm i}^2/(3t)=50k^2/(3t)$, the horizon entry time $t_{\rm H}$ is given by 
\begin{equation}
  t_{\rm H}=100\sqrt{6}/k=10\sqrt{6}L. 
\end{equation}

To clarify the spacetime structure depending on the value of $\mu$, 
we investigate the sequence of the future-directed radial-outward null geodesics. 
From the null condition, we obtain
\begin{equation}
  -\dot t^2+\frac{(\partial_r R)^2}{1-\tilde kr^2}\dot r^2=0, 
\end{equation}
where the dot ``$\dot ~$ '' denotes differentiation with respect to an affine parameter. 
The null geodesic equation is given by 
% \begin{equation}
%   \ddot t+\frac{\del_r R\del_t\del_r R}{1-\tilde kr^2}\dot r^2=0.
% \end{equation}
\begin{equation}
  \ddot t+\frac{(\del_r R)\del_t\del_r R}{1-\tilde kr^2}\dot r^2=0.
\end{equation}

Now we are interested in whether the spacetime is globally naked or locally naked~\cite{Eardley:1978tr}. 
The spacetime singularity is located at $R=0$, indicating 
$t=0$ in the past and 
% \AEE{I think the expoent on k should be negative, i mean $k^{-3/2}$}
% \CY{right corrected, and k's in the profile and LTB should be distinguished.}
$t=t_{\rm s}(r):=\frac{\pi}{3}m\tilde k^{-3/2}$ after the time evolution. 
Since it can be shown that any future-directed outgoing geodesics cannot be emanated from the singularity $t=t_{\rm s}$ with $r\neq0$ (see, e.g., Ref.~\cite{Plebanski:2024vid}), 
the singularity can be naked only at $(t,r)=(t_{\rm s}(0),0)$. 
In general, multiple future-directed radial null geodesics can be emanated from this point. 
Therefore, we need to search for the most past null geodesic emanated from the point $r=0$ and $t=t_{\rm s}(0)$. 
In practice, we solve the past-directed inward null geodesic equation from the radius $(t,r)=(t_{\rm t}, 1/k)$ with $t_{\rm t}$ being the trial value, and find the critical time $t_{\rm cr}$ above which the null geodesic hits the singularity. 
Then we solve the future-directed outgoing null geodesic equation 
from $(t,r)=(t_{\rm cr}, 1/k)$. 
Since the current density profile is a compensated profile, namely, the central over-dense region is surrounded by an under-dense region, once the outgoing null geodesic reaches the under-dense region, we determine the spacetime is globally naked; otherwise, the null geodesic is terminated at the singularity, and it is locally naked. 
We may safely say PBH forms if the singularity is locally naked 
but not globally naked \cite{1981SvA....25..406P,Kokubu:2018fxy}. 

%%%%%%%%%%%%%%%%%%%%%%%%%%%<<start figure>>%%%%%%%%%%%%%%%%%%%%%%%%%%
\begin{figure}[htbp]
\begin{center}
\includegraphics[scale=1.]{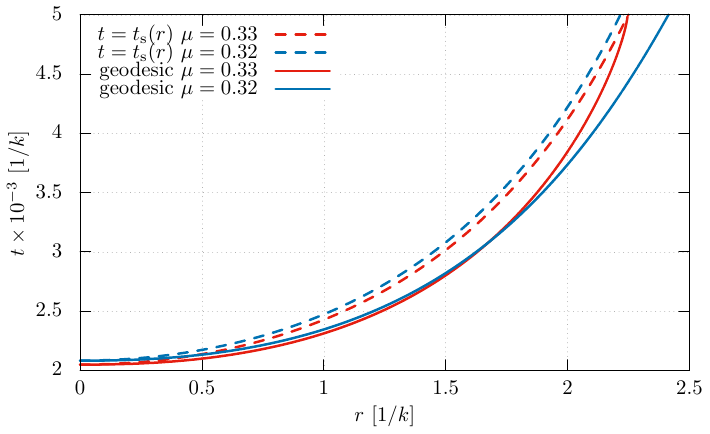}
\caption{    \baselineskip5mm 
The trajectories of the singularity $t=t_{\rm s}(r)$ and the most past null geodesic emanated from $(t,r)=(t_{\rm s}(0),0)$. 
}
\label{fig:singularity}
\end{center}
\end{figure}
%%%%%%%%%%%%%%%%%%%%%%%%%%%%<<end figure>>%%%%%%%%%%%%%%%%%%%%%%%%%%%
In Fig. \ref{fig:singularity}, we show the trajectories of the singularity $t=t_{\rm s}(r)$ (dashed line) and the most past null geodesic emanating from $(t,r)=(t_{\rm s}(0),0)$ (solid line) for $\mu=0.32$ (blue) and $\mu=0.33$ (red). 
Since the solid and dashed red lines have two intersection points, the most past null geodesic emanating from $(t,r)=(t_{\rm s}(0),0)$ terminates at the singularity. Therefore, the spacetime structure is locally naked, and the singularity is hidden by the event horizon from the asymptotic observers for $\mu=0.33$. 
On the other hand, since there are null geodesics emanating from $(t,r)=(t_{\rm s}(0),0)$ and can reach the asymptotic infinity for $\mu=0.32$, the spacetime structure is globally naked.

\subsection{Numerical simulation of spherical cases with dust fluid}

\begin{figure}[htbp]
\begin{center}
\includegraphics[scale=0.25]{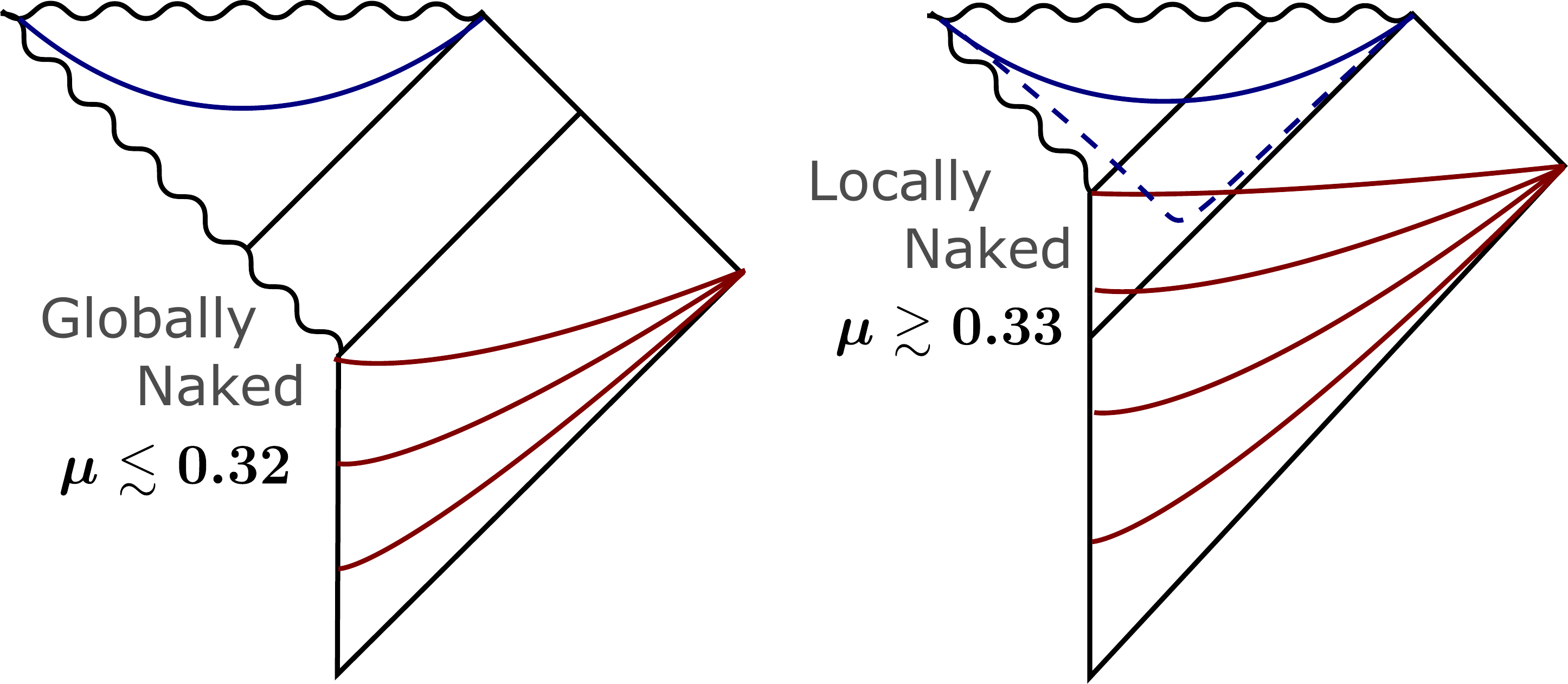}
\caption{    \baselineskip5mm 
Schematic figures for a globally naked spacetime (left) and a locally naked (right) spacetime. 
Blue curves describe the trajectories of apparent horizons. 
In the right figure, if the apparent horizon trajectory is given by the blue solid/dashed curve, 
the time slice sequence reaches the singularity/horizon before touching the horizon/singularity.  
}
\label{fig:naked}
\end{center}
\end{figure}

\begin{figure}[htbp]
\begin{center}
\includegraphics[scale=0.67]{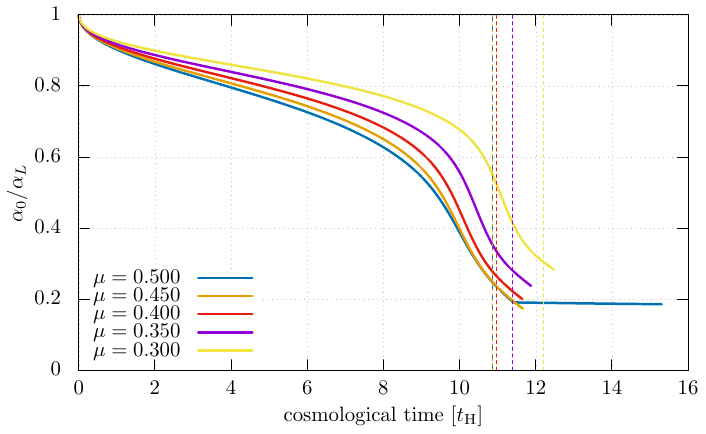}
\includegraphics[scale=0.67]{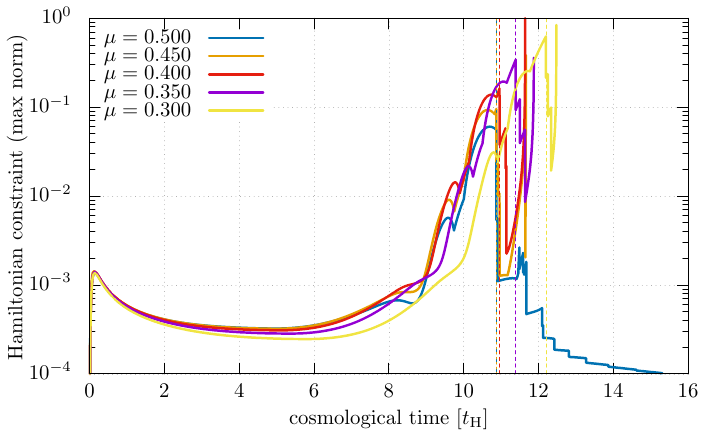}
\caption{    \baselineskip5mm 
The values of $\alpha_0/\alpha_L$ and the max norm of the Hamiltonian constraint violation 
as a function of cosmological time. The dashed vertical lines indicate the times when apparent horizons are found. 
}
\label{fig:alcon_sph}
\end{center}
\end{figure}

Although we are mainly interested in aspherical collapse, as a test bed, let us show the results of a 3+1 dimensional simulation 
for spherically symmetric cases. 
The simulations are performed in the region $0\leq X\leq L$, $0\leq Y\leq L$ and $0\leq Z\leq L$ with $X$, $Y$ and $Z$ being the Cartesian coordinates. 
We use 80 grids in each direction. 
The formation of the apparent horizon is monitored during the simulation. 
It should be noted that, since the typical spacetime structures are shown as Fig.~\ref{fig:naked}, 
even for the locally naked case, the sequence of the time slices touches the singularity before it reaches the apparent horizon. 
%%%%%%%%%%%%%%%%%%%%%%%%%%%<<start figure>>%%%%%%%%%%%%%%%%%%%%%%%%%%
%%%%%%%%%%%%%%%%%%%%%%%%%%%%<<end figure>>%%%%%%%%%%%%%%%%%%%%%%%%%%%
Therefore, even if the simulation breaks down before the apparent horizon is found, 
we cannot identify that spacetime is globally naked. 
In contrast, if an apparent horizon is found before the simulation breaks down due to the central singularity 
with an acceptable extent of constraint violation, 
since the apparent horizon is inside the event horizon, we can conclude that the spacetime is locally naked. 
In the left panel of Fig.~\ref{fig:alcon_sph}, the time evolution of the lapse function 
at the origin $\alpha_0=\alpha|_{x=y=z=0}$ normalized by the lapse function at $x=y=z=L$ ($\alpha_L$) is shown.  
The value of the Hamiltonian constraint violation in max-norm is also shown in the right panel. 
%%%%%%%%%%%%%%%%%%%%%%%%%%%<<start figure>>%%%%%%%%%%%%%%%%%%%%%%%%%%
%%%%%%%%%%%%%%%%%%%%%%%%%%%%<<end figure>>%%%%%%%%%%%%%%%%%%%%%%%%%%%
The vertical dashed lines indicate the time when an apparent horizon is found. 
Since the constraint violation is monitored only outside the apparent horizon, 
the value of the constraint violation decreases discontinuously at the time when an apparent horizon is found. 
For the $\mu=0.5$ case, after the apparent horizon formation, the size of the horizon becomes sufficiently large, and the excision inside the horizon is performed. 
For the case $\mu=0.3$, in which the singularity is globally naked, an apparent horizon is found in relatively late times. 
At the formation time, the size is too small, and 
the constraint is significantly violated outside the apparent horizon. 
This behavior may reflect the global nakedness of the spacetime, although the difference from the cases 
which have slightly larger initial amplitudes is not distinct. 

A lesson from the test bed of the spherical cases is that, 
from numerical simulation, we may be able to find a sufficient condition for PBH formation 
finding stable time evolution even after the apparent horizon formation. 
The actual threshold amplitude for PBH formation is somewhat smaller than the value of the sufficient condition, and 
the simulation is expected to break down soon after the apparent horizon formation 
for the initial amplitude around the threshold value. 

%%%%%%%%%%%%%%%%%%%%%%%%%%%%%%%%%%%%%%%%%%%%%%%%%%%%%%%%%%%%%%%%
\section{Non-spherical Collapse}
\label{sec:asph}
%%%%%%%%%%%%%%%%%%%%%%%%%%%%%%%%%%%%%%%%%%%%%%%%%%%%%%%%%%%%%%%%
% \subsection{Profile of the curvature perturbation}
We consider the following non-spherical perturbation profile:
\begin{equation}
  \ln\Psi=\frac{\mu}{2}\exp\left[-\frac{1}{6}k^2r^2\right]\left[1+\frac{k^2}{6}\left(p(2X^2-Y^2-Z^2)+3e(Y^2-Z^2)\right)\right],
\end{equation}
where $r^{2}=X^{2}+Y^{2}+Z^{2}$ and the ellipticity $e$ and prolateness $p$ characterize the non-spherical symmetry. 
This profile is realized as the typical profile~\cite{Bardeen:1985tr} for the power spectrum: 
\begin{equation}
\mathcal P(\tilde k)\propto\frac{3\sqrt{6}}{\sqrt{\pi}}\frac{\tilde k^3}{k^3}\exp\left(-\frac{3}{2}\frac{\tilde k^2}{k^2}\right), 
\end{equation}
and reduces to Eq.~\eqref{eq:Psi} for $e=p=0$. 

\subsection{Criterion from the Zel'dovich approximation and the hoop conjecture}
First, for comparison with the numerical results, we briefly review the criterion estimated from the Zel'dovich approximation and the hoop conjecture~\cite{Thorne1972} following Ref.~\cite{Harada:2016mhb}. 
Let us begin by expressing the relation between the Eulerian coordinates $\bm X$ and the Lagrangian coordinates $\bm q$ by using the displacement vector $\bm D$ as follows:
\begin{equation}
  \bm X =\bm q +\bm D(t,\bm q). 
\end{equation}
From the mass conservation for the volume element $\dd^3 \bm q$, we obtain 
\begin{equation}
  \rho_{\rm b}\dd^3 \bm q=\rho_{\rm b}(1+\delta)\dd^3 \bm X=\rho_{\rm b}(1+\delta)\det\left(\delta_{ij}+\frac{\del D_i}{\del q_j}\right)\dd^3\bm q.  
\end{equation} 
% \AEE{I think there is some sign inconsitency in the chain of eqs(5.6) to (5.10) no?eq.(5.6) should be a + sign in the last eqaution isntead of (-), according to the deifniiton appendix in A.24, I think. Then 5.7 should be also +, am I right? also eq.(5.10) should be also +ln Psi then, right?}\CY{Right. Corrected.}
Therefore we obtain 
% \KK{I think this signature should be positive.}\CY{This must be negative. Albert pointed out the wrong sign in Eq.(5.6), right?}
\begin{equation}
  \delta=\frac{1}{\det\left(\delta_{ij}+\frac{\del D_i}{\del q_j}\right)}-1\simeq-\nabla\cdot\bm D, 
\end{equation}
where we linearized the expression on the rightmost side. 
For a self-contained explanation, we present the set of linearized equations introducing the independent scalar perturbation variables (e.g., $h$, $\xi$ and $V$) and some useful gauge-invariant variables (e.g., $\Psi_{\rm lin}$) in Appendix~\ref{sec:linear}. 
In the linearized equations in the synchronous comoving gauge, 
we obtain 
\begin{equation}
  \triangle \Psi_{\rm lin}=-4\pi \rho_{\rm b}a^2\delta=4\pi \rho_{\rm b}a^2\nabla\cdot\bm D.  
\end{equation}
This equation can be solved for $\bm D$ as follows:
\begin{equation}
  \bm D(t,\bm q)=\frac{2}{3}\frac{1}{a^2H^2}\nabla\Psi_{\rm lin}. 
\end{equation}
In the long-wavelength limit, we obtain
\begin{equation}
  \Psi_{\rm lin}\simeq 2\xi-\frac{1}{2}a^2H\del_t h. 
\end{equation}
Since $\del_t \Psi_{\rm lin}=\del_t \Phi=0$ for growing modes, 
for the synchronous comoving gauge, we find 
\begin{equation}
  \del_t h=\frac{2}{a}V=\frac{4}{3}\frac{1}{a^2H}\Psi_{\rm lin}, 
\end{equation}
where 
we have used Eqs.~\eqref{eq:V}, \eqref{eq:PsiPhi} and \eqref{eq:VPhi}. 
Then we obtain 
\begin{equation}
\frac{5}{6}  \Psi_{\rm lin}\simeq \xi\simeq \Psi-1\simeq \ln\Psi. 
\end{equation}
That is, 
\begin{equation}
  \bm D\simeq \frac{4}{5}\frac{1}{a^2H^2}\nabla\ln\Psi. 
\end{equation}

\begin{figure}[h!]
\begin{center}
\includegraphics[scale=1.]{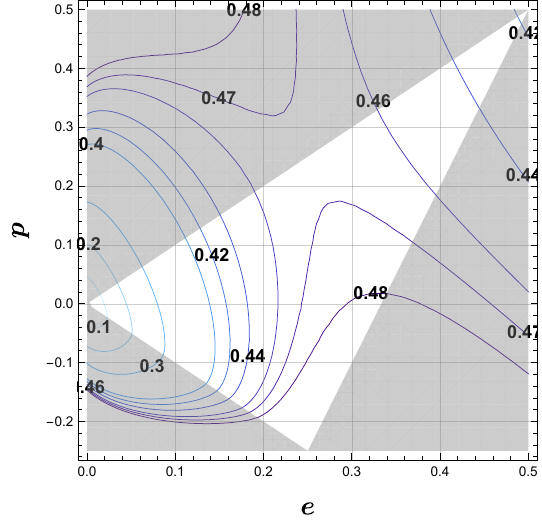}
\caption{    \baselineskip5mm 
Contour map of $\hat h(e,p)$. The shaded region is excluded by the condition $\tilde \alpha\geq\tilde \beta\geq\tilde \gamma\geq0$. 
}
\label{fig:hcont}
\end{center}
\end{figure}

Following Ref.~\cite{Harada:2016mhb} (see also Ref.~\cite{Saito:2024hlj}), 
let us introduce the following function: 
\begin{equation}
  h(\tilde \alpha,\tilde \beta,\tilde \gamma):=\frac{2}{\pi}\frac{\tilde \alpha-\tilde \gamma}{\tilde \alpha^2}E\left(\sqrt{1-\left(\frac{\tilde \alpha-\tilde \beta}{\tilde \alpha-\tilde \gamma}\right)^2}\right), 
  \label{eq:h}
\end{equation}
where $\tilde \alpha$, $\tilde \beta$ and $\tilde \gamma$ are the eigenvalues of $\left.-\nabla \bm D\right|_{r=0}$ with $\tilde \alpha\geq\tilde \beta\geq\tilde \gamma$. 
In Ref.~\cite{Harada:2016mhb}, from the hoop conjecture~\cite{Thorne1972}, 
the criterion for PBH formation is given by $h(\tilde\alpha,\tilde \beta,\tilde \gamma)<1$ at the horizon entry. 
For our specific form of $\Psi$, we have 
\begin{equation}
  \left(\tilde \alpha,\tilde \beta,\tilde \gamma\right)
  =\frac{2\mu}{15}\frac{k^2}{a^2H^2}\left(1+3e+p,1-2p,1-3e+p\right). 
\end{equation}
Substituting these expressions into Eq.~\eqref{eq:h}, the criterion for PBH formation is given by
\begin{equation}
  \label{eq:ana_cri}
  \mu>\hat h(e,p)\times \frac{6a^2H^2}{k^2}
  % \mu\frac{k^2}{a^2H^2}h=\hat h(e,p):=\frac{90}{\pi}\frac{e}{(1+3e+p)^2}E\left(\sqrt{1-\frac{(e+p)^2}{4e^2}}\right). 
\end{equation}
with 
\begin{equation}
  \hat h(e,p):=\frac{15}{\pi}\frac{e}{(1+3e+p)^2}E\left(\sqrt{1-\frac{(e+p)^2}{4e^2}}\right). 
\end{equation}
The contour map of $\hat h(e,p)$ is depicted in Fig.~\ref{fig:hcont}. 
%%%%%%%%%%%%%%%%%%%%%%%%%%%<<start figure>>%%%%%%%%%%%%%%%%%%%%%%%%%%
%%%%%%%%%%%%%%%%%%%%%%%%%%%%<<end figure>>%%%%%%%%%%%%%%%%%%%%%%%%%%%

% %%%%%%%%%%%%%%%%%%%%%%%%%%%%%%%%%%%%%%%%%%%%%%%%%%%%%%%%%%%%%%%%
% \subsection{Initial Data Setting}
% %%%%%%%%%%%%%%%%%%%%%%%%%%%%%%%%%%%%%%%%%%%%%%%%%%%%%%%%%%%%%%%%

%%%%%%%%%%%%%%%%%%%%%%%%%%%%%%%%%%%%%%%%%%%%%%%%%%%%%%%%%%%%%%%%
\subsection{Comparison with numerical simulation}
\begin{figure}[htbp]
\begin{center}
\includegraphics[scale=0.67]{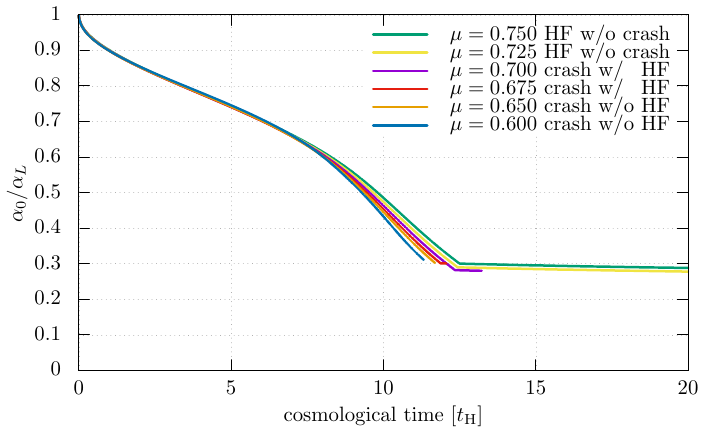}
\includegraphics[scale=0.67]{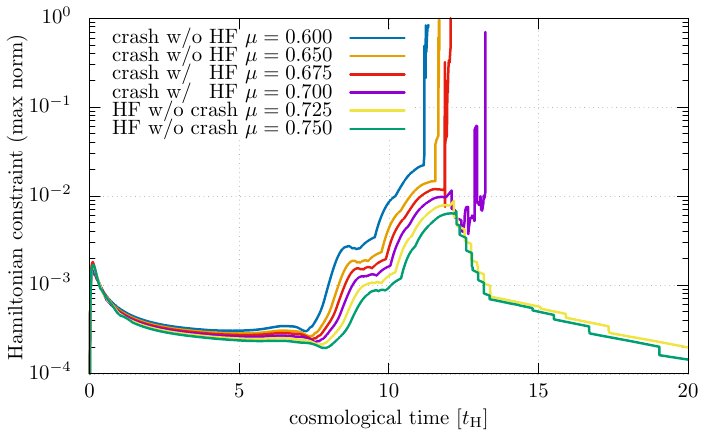}
\caption{    \baselineskip5mm 
The values of $\alpha_0/\alpha_L$ and the max norm of the Hamiltonian constraint violation 
as a function of the cosmological time for each value of the initial amplitude $\mu$ with $p=0$ and $e=0.1$. 
The legend labels ``HF w/o crash", ``crash w/ HF", and ``crash w/o HF" indicate 
stable evolution with horizon formation, crash after horizon formation, and crash without horizon formation, respectively. 
}
\label{fig:alcon_e01}
\end{center}
\end{figure}

%%%%%%%%%%%%%%%%%%%%%%%%%%%%%%%%%%%%%%%%%%%%%%%%%%%%%%%%%%%%%%%%
Here, let us focus on the $p=0$ cases. 
We use 80 grids for each direction, as with the spherically symmetric cases. 
In the left panel of Fig.~\ref{fig:alcon_e01}, the time evolution of the value of $\alpha_0/\alpha_L$ is shown.  
The value of the Hamiltonian constraint violation in the max norm is also shown in the right panel. 
%%%%%%%%%%%%%%%%%%%%%%%%%%%<<start figure>>%%%%%%%%%%%%%%%%%%%%%%%%%%
%%%%%%%%%%%%%%%%%%%%%%%%%%%%<<end figure>>%%%%%%%%%%%%%%%%%%%%%%%%%%%
The results can be classified into three cases: stable evolution with horizon formation (``HF w/o crash"), 
crash after horizon formation (``crash w/ HF"), and crash without horizon formation (``crash w/o HF"). 
The smallest value of $\mu$ in the case ``HF w/o crash" may give a sufficient condition for 
PBH formation. The real threshold may exist below the value giving the sufficient condition. 

\begin{figure}[htbp]
\begin{center}
\includegraphics[scale=1.5]{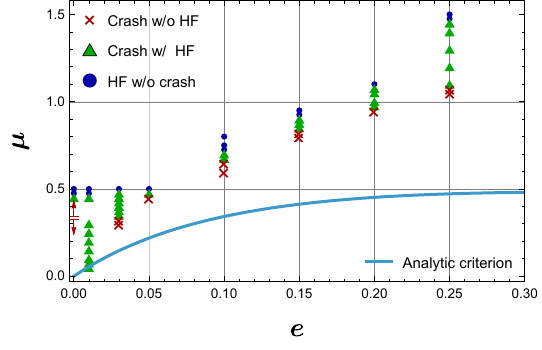}
\caption{    \baselineskip5mm 
% \AEE{In the figure, for e=0, we would expect that muc goes to zero right?, maybe can be said some comment expliclty, I think i didnt see such comment}
Classification of the time evolutions characterized by the initial amplitude $\mu$ and the ellipticity $e$. 
The blue solid line describes the line $\mu=\hat h(e,0)$. 
The red arrows indicate the threshold between the globally naked (downward) and locally naked (upward) cases of the LTB solution. }
% \CY{Because of the shell focusing singularity discussed in Sec.IV, muc does not go to zero if we take the effect of inhomogeneity into account. Anyway, I will add some comment somewhere. $\rightarrow$added. }
\label{fig:thre_p0}
\end{center}
\end{figure}

Let us compare our results of numerical simulation for several values of $e$ with $p=0$. 
We perform the numerical simulations to find the classification of the cases for $e=0$, $0.01$, $0.03$, $0.05$, $0.1$, $0.15$, $0.2$, and $0.25$. 
In Fig.~\ref{fig:thre_p0}, we show the classifications obtained from numerical simulations in the $\mu$-$e$ plane. 
%%%%%%%%%%%%%%%%%%%%%%%%%%%<<start figure>>%%%%%%%%%%%%%%%%%%%%%%%%%%
%%%%%%%%%%%%%%%%%%%%%%%%%%%%<<end figure>>%%%%%%%%%%%%%%%%%%%%%%%%%%%
According to the above discussion, the lower limit of the blue circle (``HF w/o crash") may give a sufficient 
condition for PBH formation. 
It should be noted that
the 
effect of the inhomogeneity discussed in Sec.~\ref{sec:sph} is not taken into account in the 
line of the analytic criterion with $aH=k/\sqrt{6}$  
(blue solid line in Fig.~\ref{fig:thre_p0}). 
Therefore, although the line of the analytic criterion approaches $\mu=0$ at $e=0$, 
black hole formation is not necessarily expected 
even for $e=0$. 
Although we cannot give a lower bound of the threshold for $\mu$, since the analytic estimation with $aH=k/\sqrt{6}$ gives obviously smaller values compared to the boundary between the green triangles (``crash w/ HF") and the red crosses (``crash w/o HF"), the real threshold might be larger than the values indicated by the blue solid line. 

Our results suggest that the threshold value might be significantly larger, so PBH formation would be harder than the analytic estimation within the range in which the fluid approximation is valid. 
However, the failure of the fluid description associated with the singular behavior does not necessarily mean the end of the gravitational collapse of more physically realistic matter fields. 
If we may accept a subsequent time evolution avoiding the serious curvature singularity, 
there is a possibility for the system to eventually collapse into a black hole. 
In the following section, we will describe 
a suggestive example of numerical simulation. As one possible extension beyond the single-stream dust approximation, we assume collisionless matter for simplicity. 

%%%%%%%%%%%%%%%%%%%%%%%%%%%%%%%%%%%%%%%%%%%%%%%%%%%%%%%%%%%%%%%%
\section{A collisionless particle system}
\label{sec:cps}
%%%%%%%%%%%%%%%%%%%%%%%%%%%%%%%%%%%%%%%%%%%%%%%%%%%%%%%%%%%%%%%%

\subsection{A collisionless particle system and stress-energy tensor}

Let us consider the collisionless particle system composed of 
$N$ particles, each of which travels a timelike geodesic. 
The four-velocity $u^\mu$ of a particle can be decomposed 
as follows\cite{Vincent:2012kn}:
\begin{equation}
u^\mu=\Gamma\left(n^\mu+V^\mu\right), 
\end{equation}
where the spatial velocity components $V^\mu$ satisfy  
$V^\mu n_\mu=0$. 
Then, the 3+1 decomposition of geodesic equations is 
expressed as follows\cite{Vincent:2012kn}:
\begin{eqnarray}
\frac{\dd \Gamma}{\dd t}&=&\Gamma V^i\left(\alpha K_{ij} V^j-\del_i \alpha\right), \\
\frac{\dd V^i}{\dd t}&=&\alpha V^j 
\left[V^i \left(\del_j\ln \alpha-K_{jk}V^k\right)
+2K^i_{~j}-V^k\Gamma^i_{jk}\right]
-\gamma^{ij}\del_j\alpha-V^j\del_j\beta^i, \\
\frac{\dd \tau}{\dd t}&=&\frac{\alpha}{\Gamma},
\end{eqnarray}
% \AEE{I think in Eq.6.3 for left side shoudl be $dV^{i}/dt$, there is free index i} \CY{right. corrected} 
where $\tau$ is the proper time and $\dd/\dd t=\frac{\alpha}{\Gamma}u^\mu \del_\mu$. 

The energy-momentum tensor for a particle system 
is given by (see, e.g., \cite{Misner:1974qy})
\begin{equation}
T^{\mu\nu}=-\sum_p m_p
\frac{\delta^3\left(\bm x-\bm x_p\right)}{u^\lambda_p n_\lambda \sqrt{\gamma}}
u^\mu_pu^\nu_p, 
\end{equation}
where $m_p$ is the proper mass of the particle labeled by $p$ and $\bm x$ and $\bm x_p$ denote 
the spatial coordinates and those values at the particle position, respectively. 
Eqs.~(\ref{eq:rhon}--\ref{eq:sij}) are reduced to 
\begin{eqnarray}
E&=&
\sum_p m_p \Gamma_p\frac{\delta^3\left(\bm x -\bm x_p\right)}{\sqrt{\gamma}},
\label{eq:rhon2}
\\
J^i&=&\sum_p m_p \Gamma_p V^i_p \frac{\delta^3\left(\bm x-\bm x_p\right)}{\sqrt{\gamma}},
\label{eq:ji2}
\\
S^{ij}&=&\sum_pm_p \Gamma_p V^i_p V^j_p 
\frac{\delta^3\left(\bm x-\bm x_p\right)}{\sqrt{\gamma}}. 
\label{eq:sij2}
\end{eqnarray}

Since the delta functions cannot be treated numerically, for each particle, 
we instead assign a cubic domain based on the coordinate spacing and 
consider uniform distributions inside the cubic box. 
Namely, the delta function is replaced by the uniform support function as 
\begin{equation}
  \delta^3(\bm x-\bm x_p)\rightarrow\Theta\left(\frac{\Delta x}{2}-|x-x_p|\right)\Theta\left(\frac{\Delta y}{2}-|y-y_p|\right)\Theta\left(\frac{\Delta z}{2}-|z-z_p|\right)\frac{1}{\Delta x\Delta y\Delta z}, 
\end{equation}
and $\sqrt{\gamma}$ is evaluated at $\bm x=\bm x_p$.

\subsection{Particle settings}
In order to convert the fluid distribution into a particle distribution, 
we consider a regularly aligned initial distribution of particles. 
In addition, we assume the numerical region is initially filled with no space and no overlap between particles. 
Since the particles are properly aligned initially, 
we cannot introduce inhomogeneity by an inhomogeneous particle distribution. 
Instead of an inhomogeneous distribution, we introduced different proper masses for individual particles so that the density distribution may have the desired inhomogeneity. 
That is, we set each particle mass $m_p$ as 
\begin{equation}
  m_p=\frac{E(\bm x_p)}{\Gamma_p}\sqrt{\gamma(\bm x_p)}\Delta x\Delta y\Delta z 
\end{equation}
on the initial hypersurface. 

\subsection{Results of the numerical simulation}
First, as is described in Appendix~\ref{sec:paandltb}, 
we performed the comparison between the numerical simulation of 
the spherical case and the corresponding LTB solution for $\mu=1.3$. 
We can find good agreement for this parameter setting until the time of horizon formation. 

\begin{figure}[htbp]
\begin{center}
\includegraphics[scale=1.]{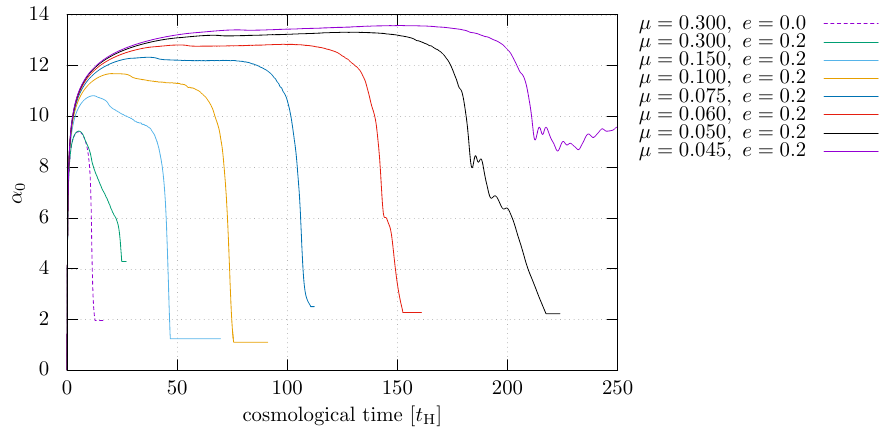}
\caption{    \baselineskip5mm 
Time evolution of $\alpha_0$ for each value of $\mu$ with $e=0.2$ and $N=80$. 
The case for $e=0$ and $\mu=0.3$ is also plotted as a reference.}
\label{fig:alpha_p}
\end{center}
\end{figure}

In this section, we are mainly interested in the situation where 
the simulation breaks down for the dust fluid case. 
As an example, let us focus on the case $e=0.2$. 
According to Fig.~\ref{fig:thre_p0}, the calculation breaks down before horizon formation 
for $\mu\leq0.95$. 
However, for the particle simulation, the calculation does not crash even for much smaller values of $\mu$. 
In Fig.~\ref{fig:alpha_p}, we show the value of $\alpha_0$ as a function of the cosmological time for each value of $\mu$. 
We also plot the case for $e=0$ and $\mu=0.3$ as a reference. 
%%%%%%%%%%%%%%%%%%%%%%%%%%%<<start figure>>%%%%%%%%%%%%%%%%%%%%%%%%%%
%%%%%%%%%%%%%%%%%%%%%%%%%%%%<<end figure>>%%%%%%%%%%%%%%%%%%%%%%%%%%%
As is shown in Fig.~\ref{fig:alpha_p}, the value of $\alpha_0$ starts to steeply decrease from a certain time and 
an apparent horizon is formed for $\mu\geq0.05$. 
In contrast, for $\mu\leq0.045$, the decrease of $\alpha_0$ stops, and it keeps a middle value without horizon formation. 

\begin{figure}[htbp]
\begin{center}
\includegraphics[scale=0.6]{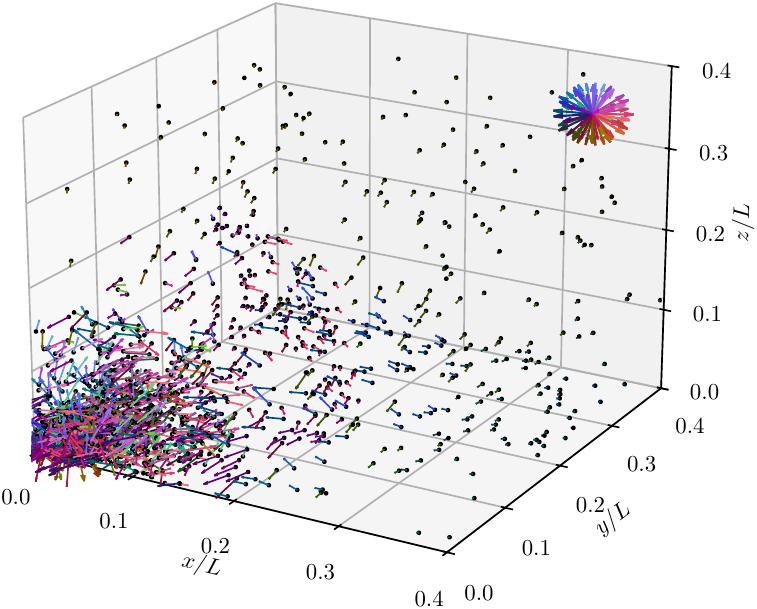}
\includegraphics[scale=0.55]{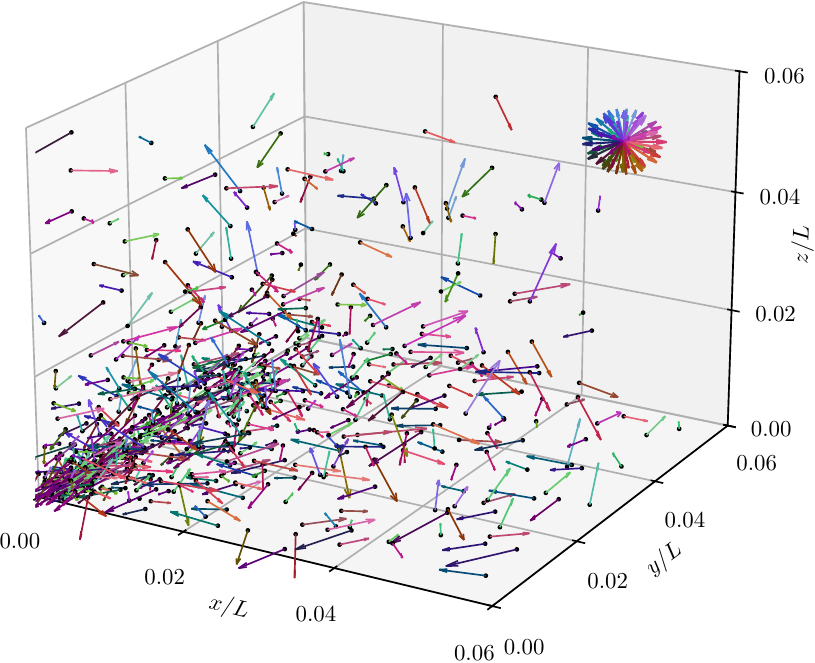}
\caption{    \baselineskip5mm 
Particle distribution at $t=245t_{\rm H}$. 
The panel on the right is an enlargement of the central part of the figure on the left.  The number of particles has been reduced to 1/80.} 
% \AEE{i think very hard to see no?...maybe that options could look better: 1)make 2D surface plot instead of 3d. 2) still using 3D plot but put different background colours to the data points, to show for instance velocities of the particles with a colour legend. 3)make some histogram of positions and number particles instead of representing all number particles in 3D plot}\CY{how about it?}
\label{fig:veldis}
\end{center}
\end{figure}

Let us check the overall dynamics following snapshots of the density profile. 
%%%%%%%%%%%%%%%%%%%%%%%%%%%<<start figure>>%%%%%%%%%%%%%%%%%%%%%%%%%%
\begin{figure}[htbp]
\begin{center}
\includegraphics[scale=0.6]{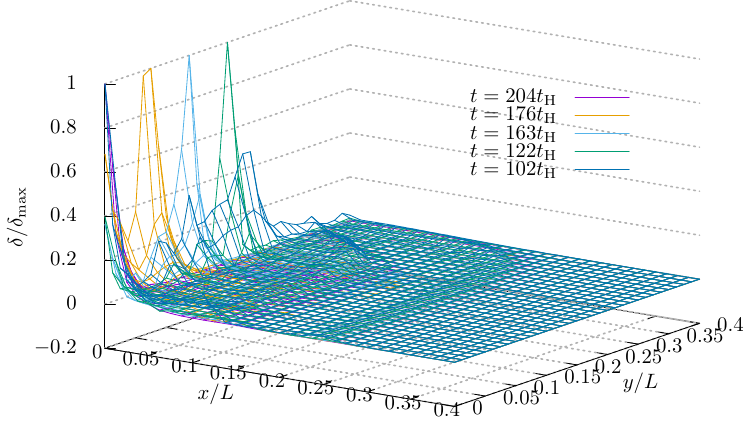}
\includegraphics[scale=0.6]{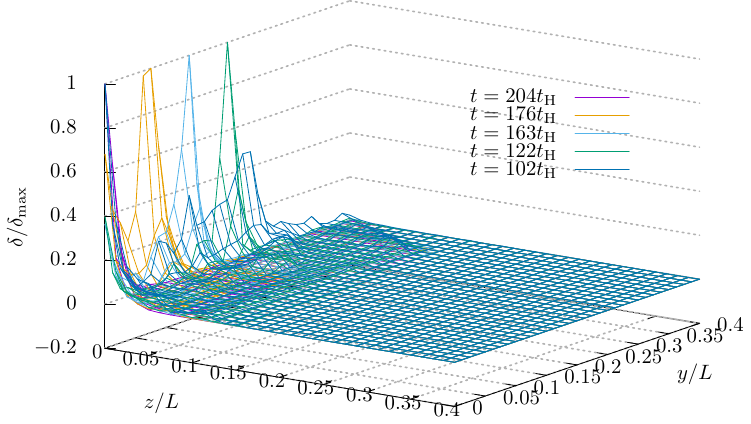}
\includegraphics[scale=0.6]{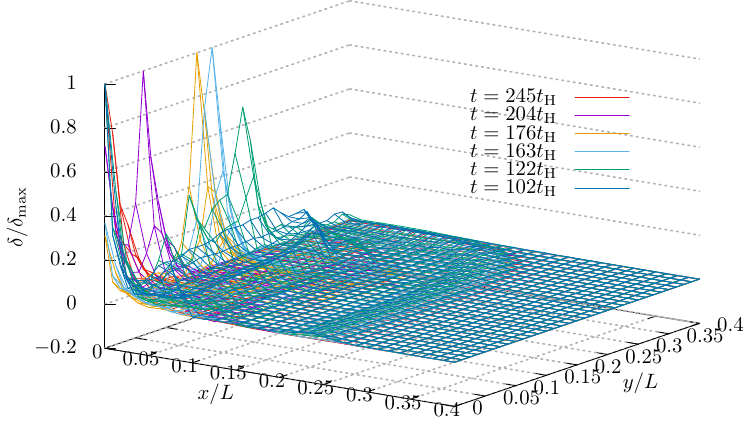}
\includegraphics[scale=0.6]{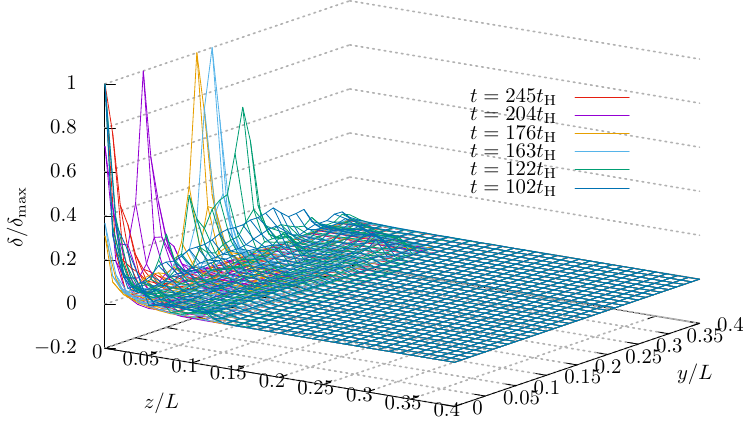}
\caption{    \baselineskip5mm 
Snapshots of the profiles of the density perturbation normalized by its maximum value $\delta_{\rm max}$ for $\mu=0.05$ (upper) and $0.045$ (lower) on the $x$-$y$ plane (left) and $y$-$z$ plane (right).}
\label{fig:density}
\end{center}
\end{figure}
%%%%%%%%%%%%%%%%%%%%%%%%%%%%<<end figure>>%%%%%%%%%%%%%%%%%%%%%%%%%%%
It can be found that the density perturbation $\delta=8\pi E/(3 H^2)-1$
first increases around the center and the tip of the over-dense distribution along the longest distribution axis, which is the $y$-direction in the present case. 
This behavior around the tip of the over-dense region is similar to the behavior reported in Refs.~\cite{Shapiro:1991zza,Yoo:2016kzu}. 
In the case $\mu=0.045$, the collapse halts, and it may describe a halo formation below the threshold amplitude. 
This halo would be supported by the effective pressure generated by the velocity dispersion of the particles. 
To explicitly see that, we plot the particle distribution in Fig.~\ref{fig:veldis} together with the arrows presenting the velocity of each particle. 
%%%%%%%%%%%%%%%%%%%%%%%%%%%<<start figure>>%%%%%%%%%%%%%%%%%%%%%%%%%%
%%%%%%%%%%%%%%%%%%%%%%%%%%%%<<end figure>>%%%%%%%%%%%%%%%%%%%%%%%%%%%
One can find that the velocities of the particles are not coherent in the central region, and velocity dispersion is associated with the random motion.

\begin{figure}[h!]
\begin{center}
\includegraphics[scale=1.]{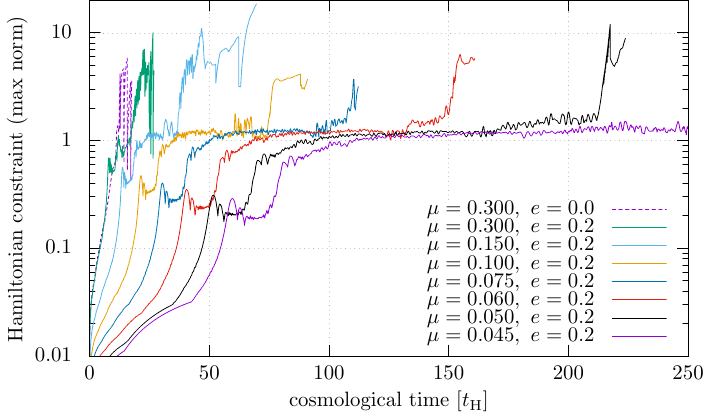}
\includegraphics[scale=1.]{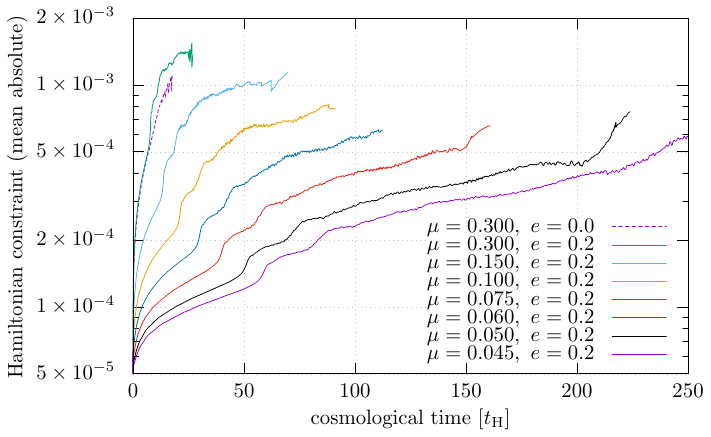}
\caption{    \baselineskip5mm 
Time evolution of the max norm (upper panel) and the mean absolute value (lower panel) of the Hamiltonian constraint violation for each value of $\mu$ with $e=0.2$. 
% \AEE{one question, the maximum of the constraint where it happens? is hihgly localized near the AH region for instance? or like near boundaries etc(then may be safe if is the case, but provably not).}
}
\label{fig:constr_p}
\end{center}
\end{figure}

\begin{figure}[h!]
\begin{center}
\includegraphics[scale=0.6]{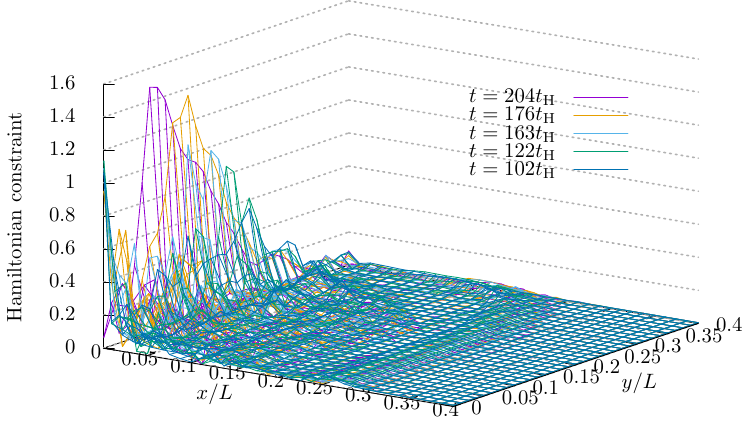}
\includegraphics[scale=0.6]{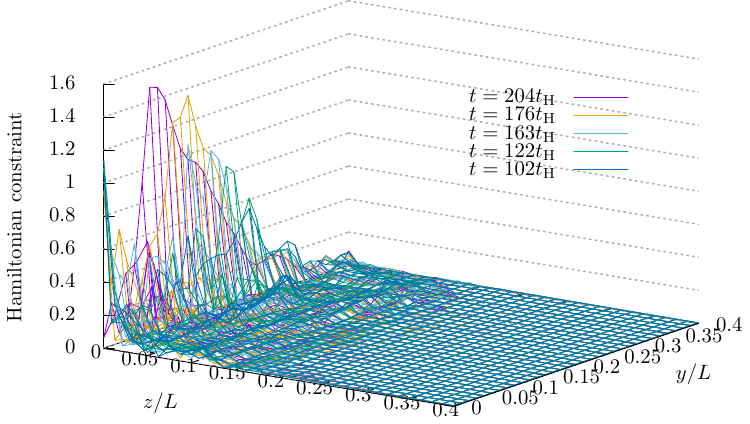}
\includegraphics[scale=0.6]{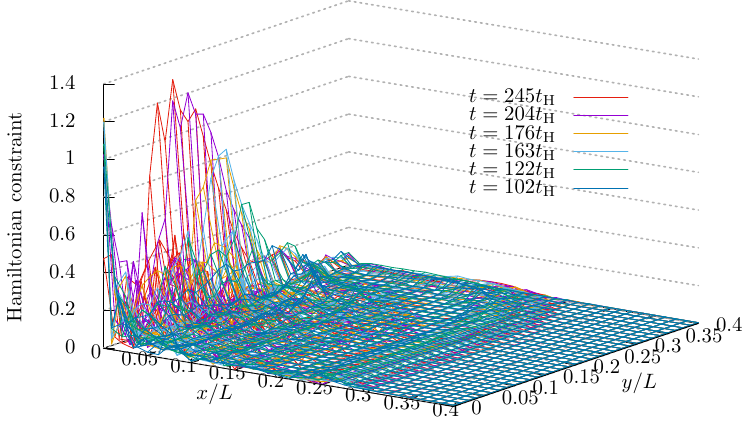}
\includegraphics[scale=0.6]{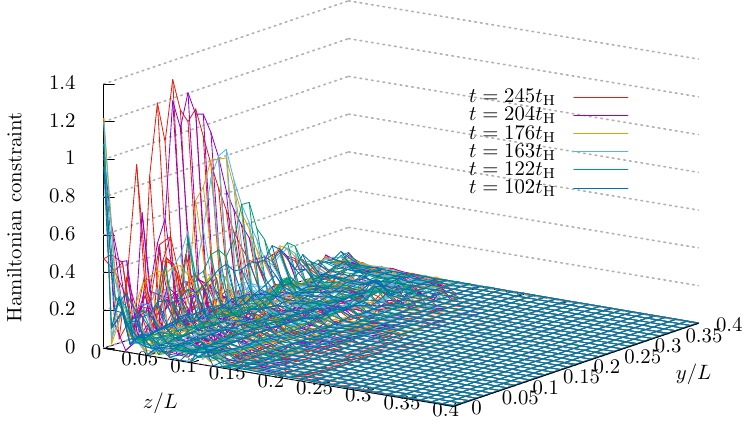}
\caption{    \baselineskip5mm 
Snapshots of the profiles of the Hamiltonian constraint violation for $\mu=0.05$ (upper) and $0.045$ (lower) on the $x$-$y$ plane (left) and $y$-$z$ plane (right).}
\label{fig:ham}
\end{center}
\end{figure}

The calculation does not break down for $\mu\leq0.045$ and 
will continue to run unless manually stopped. 
% continue until the termination instruction is given 
% \THc{Not clear to me. OK!}. 
If one accepts this result, the threshold of PBH formation is at around $\mu\sim 0.045$, which is much smaller than 
the values indicated by Fig.~\ref{fig:thre_p0} and even smaller than the analytic estimation \eqref{eq:ana_cri} 
with $aH=k/\sqrt{6}$. 
However, there is a big caveat. 
Although the calculation does not break down, constraints are significantly violated, as shown in Fig.~\ref{fig:constr_p} and \ref{fig:ham}. 
%%%%%%%%%%%%%%%%%%%%%%%%%%%<<start figure>>%%%%%%%%%%%%%%%%%%%%%%%%%%
%%%%%%%%%%%%%%%%%%%%%%%%%%%%<<end figure>>%%%%%%%%%%%%%%%%%%%%%%%%%%%
%%%%%%%%%%%%%%%%%%%%%%%%%%%<<start figure>>%%%%%%%%%%%%%%%%%%%%%%%%%%
%%%%%%%%%%%%%%%%%%%%%%%%%%%%<<end figure>>%%%%%%%%%%%%%%%%%%%%%%%%%%%
The origin of the localized numerical constraint violation requires further investigation and limits the quantitative interpretation of the particle results.
\footnote{
The constraint violation may be regarded as a mismatch between the stress-energy tensor described by the particle system and the geometry. 
Then we might expect that this mismatch originates from the failure of the collisionless particle description and can be resolved by introducing more physically realistic matter fields. 
The constraint violation occurs around when and where the particles intersect and cross each other. 
This situation corresponds to the shell crossing or shell focusing singularity for the dust fluid case. 
Since the matter density and curvature are divergent at the shell crossing and shell focusing singularity, 
a more realistic description of the matter field than dust fluid and particle system would be needed to properly describe the small-scale dynamics around the high-density regions. 
}
Nevertheless, it would be suggestive enough showing an example of the numerical simulation in which the mean absolute value of the Hamiltonian constraint is small enough (see the lower panel of Fig.~\ref{fig:constr_p}). 
That is, the violation of the constraint equations remains localized. 

%%%%%%%%%%%%%%%%%%%%%%%%%%%%%%%%%%%%%%%%%%%%%%%%%%%%%%%%%%%%%%%%
\section{Summary}
\label{sec:conclusion}
%%%%%%%%%%%%%%%%%%%%%%%%%%%%%%%%%%%%%%%%%%%%%%%%%%%%%%%%%%%%%%%%

We have investigated the formation of primordial black holes in an early matter-dominated universe using fully nonlinear 3+1 numerical relativity. 
We constructed initial data for the dust fluid using a long-wavelength curvature perturbation with ellipticity, 
allowing us to study both spherical and triaxial collapse configurations. 
After reviewing spherically symmetric dust collapse and the connection with long-wavelength solutions, we examined the limitations of the dust fluid description in a matter-dominated background. 
The dust fluid simulations show that sufficiently large amplitude perturbations can form apparent horizons, while smaller amplitudes fail to form a black hole before the simulation breaks down. 
The breakdown of the simulation is caused by shell focusing or shell-crossing singularities that appear in the dust fluid description.

To address this breakdown of the fluid approximation in a concrete collisionless realization, we also implemented a collisionless particle system in full numerical relativity. 
The particle-based simulations avoid the immediate crash associated with fluid shell crossing, 
and they demonstrate horizon formation for amplitudes larger than a certain value. 
The resulting collapse threshold is significantly smaller than earlier analytic estimates based on simplified matter-dominated criteria, 
although the late-time evolution is accompanied by significant Hamiltonian constraint violations after the particle intersection happens.

If one interprets the failure of the fluid approximation as indicating that singularity formation associated with shell focusing or shell crossing suppresses PBH formation, 
then the threshold may be significantly larger than analytic estimates, and PBH formation may be harder to realize. 
On the other hand, if one accepts the particle simulation results shown here, PBH formation becomes much easier than previously expected. 
Thus, our results forcefully suggest that there is a large theoretical uncertainty in the PBH formation criterion during a matter-dominated epoch.

Here, we should remark on the impeding mechanism for PBH formation in a matter-dominated era associated with black hole spin. 
Throughout the analyses in Refs.~\cite{Harada:2017fjm,Saito:2024hlj,Ye:2025wif}, the conclusion has evolved, 
but the latest work suggests that spin effects are negligibly small compared with the non-spherical collapse suppression mechanisms studied here. 
However, these spin estimates were obtained using analytic models based on the Zel'dovich approximation and the hoop conjecture, 
so the change of the threshold value in the non-spherical collapse studied numerically in this paper may also play an important role in the spin effect. 
Then continued analyses of spin effects remain necessary. 

Future work should refine the treatment of the high-density regime, improve constraint preservation, and extend the analysis to quantify the PBH abundance and the possible gravitational wave signatures associated with matter-dominated collapse. 
These developments are necessary to make definitive predictions for PBH formation in realistic early-universe scenarios.

%%%%%%%%%%%%%%%%%%%%%%%%%%%%%%%%%%%%%%%%%%%%%%%%%%%%%%%%%%%%%%%%
\section*{Acknowledgements}
%%%%%%%%%%%%%%%%%%%%%%%%%%%%%%%%%%%%%%%%%%%%%%%%%%%%%%%%%%%%%%%%
%This work was supported by JSPS KAKENHI Grant Numbers JP25K07281 (C.Y.) and JP24K07027 (C.Y.). 

This work was supported by JSPS KAKENHI Grant Numbers JP25K07281 (C.Y.), JP24K07027 (C.Y., T.H., and K.K.) and JP26K17141 (A.E.)

\appendix

\titleformat{\section}  % which section command to format
  {\fontsize{14}{16}\bfseries} % format for whole line
  {\Alph{section}.} % how to show number
  {0.5em} % space between number and text
  {} % formatting for just the text
  [] % formatting for after the text

\titleformat{\subsection}  % which section command to format
  % {\fontsize{12}{14}\sffamily} % format for whole line
  {\fontsize{12}{14}\bfseries} % format for whole line
  {\arabic{subsection}.} % how to show number
  {0.5em} % space between number and text
  {} % formatting for just the text
  [] % formatting for after the text

  % \titleformat{\subsubsection}  % which section command to format
  % {\fontsize{10}{12}\bfseries} % format for whole line
  % {\arabic{subsubsection}.} % how to show number
  % {0.5em} % space between number and text
  % {} % formatting for just the text
  % [] % formatting for after the text

% \renewcommand{\thesection}{\Alph{section}}
% \renewcommand{\thesubsection}{\arabic{subsection}}
\renewcommand{\theequation}{\thesection.\arabic{equation}}

%%%%%%%%%%%%%%%%%%%%%%%%%%%%%%%%%%%%%%%%%%%%%%%%%%%%%%%%%%%%%%%%
\section{Linear perturbation equations}
\label{sec:linear}
%%%%%%%%%%%%%%%%%%%%%%%%%%%%%%%%%%%%%%%%%%%%%%%%%%%%%%%%%%%%%%%%

In this section, we consider the linear cosmological scalar perturbation on 
a background homogeneous and isotropic spatially flat spacetime. 
We consider the perfect fluid 
with the linear equation of state $p=w\rho$, and introduce the cosmological constant $\Lambda$. 
We denote the small parameter with respect to the perturbation amplitude as $\epsilon$.  
Then, the order of each variable introduced in Sec.~\ref{sec:bssn} is given as follows:
\begin{equation}
\psi=1+\mathcal O(\epsilon),~\tilde\gamma_{ij}=f_{ij}+\mathcal O(\epsilon),
~K=-3H+ O(\epsilon),~\tilde A_{ij}= O(\epsilon),~
\rho=\rho_{\rm b}+ O(\epsilon),  
\end{equation}
where $H$ is the background Hubble function and $\rho_{\rm b}=(3H^2-\Lambda)/(8\pi)$. 
% \AEE{it is said that a cosmological constant is introduced, but then the deifnition of rhob would be $\rho_{\rm b}=(3H^2-\Lambda)/(8\pi)$ right?}\CY{Right}. 

We introduce the following perturbation variables of $\mathcal O(\epsilon)$:
\begin{eqnarray}
\xi&:=&\psi-1, \label{eq:xi}\\
\chi&:=&\alpha_{\rm c}-1, \\
\kappa&:=&-\frac{K+3H}{3H}, \\
\delta&:=&\frac{\rho-\rho_{\rm b}}{\rho_{\rm b}}, \\
h_{ij}&:=&\tilde \gamma_{ij}-f_{ij}. 
\end{eqnarray}
Since we are interested in the scalar perturbation, 
introducing the two scalar variables $h$ and $A$, 
we assume that $h_{ij}$ and $\tilde A_{ij}$ are respectively given as 
\begin{eqnarray}
h_{ij}&=&\left(\mathcal D_i\mathcal D_j -\frac{1}{3}f_{ij}\triangle\right)h, \label{eq:hij}\\
\tilde A_{ij}&=&\left(\mathcal D_i\mathcal D_j -\frac{1}{3}f_{ij}\triangle\right)A. 
\end{eqnarray}
Then, at the linear order, the equations of motion can be written as follows: 
% \AEE{i think in eq.(A.9) should be $\Lambda/3$, not $3 \Lambda$}\CY{right, . Corrected and I rearranged some expressions a bit. }
\begin{eqnarray}
\triangle \xi&=&\frac{1}{12}\triangle\triangle h -\frac{1}{4}a^2(3H^2-\Lambda)\delta+\frac{3}{2}a^2H^2\kappa, \\
\triangle A&=&-3H\kappa+\frac{3}{2}(3H^2-\Lambda)(1+w)(v+\beta), \\
\partial_t A&=&\frac{1}{a^2}\left(-2\xi+\frac{1}{6}\triangle h -\chi\right)-3HA, \\
\partial_t \xi&=&\frac{1}{2}H(\chi+\kappa)+\frac{1}{6a^2}\triangle \beta, \\
\partial_t \kappa&=&-\frac{1+w}{2H}\left(3H^2-\Lambda\right)\chi+\frac{1}{3a^2H}\triangle \chi\cr
			&&-\frac{1}{2H}\left[(1-3w)H^2+(1+w)\Lambda\right]\kappa 
			-\frac{1+3w}{6H}\left(3H^2-\Lambda\right)\delta, \\
\partial_t h&=&-2A+\frac{2}{a^2}\beta, \\
6\partial_t\xi+\partial_t \delta&=&-(1+w)\frac{1}{a^2}\triangle v-w\frac{1}{a^2}\triangle \beta -3wH(\chi+\kappa), \\
\partial_t v+\partial_t \beta&=&3wH(v+\beta)-\frac{w}{1+w}\delta-\chi. 
\end{eqnarray}

A useful set of gauge-invariant variables is given by 
\begin{eqnarray}
\Psi_{\rm lin}&:=&2\xi+H\left(\beta-\frac{1}{2}a^2\del_th\right)-\frac{1}{6}\triangle h, 
\label{eq:PsiLin}\\
\Phi&:=&\chi+\partial_t \left(\beta-\frac{1}{2}a^2\partial_t h\right), \\
\Delta&:=&\delta-3(1+w)H(\beta+v), \\
V&:=&\frac{v}{a}+\frac{1}{2}a\partial_t h, 
\label{eq:V}\\
\mathcal K&:=&\kappa-\frac{1}{H}\left[\frac{1}{2}(1+w)\left(3H^2-\Lambda\right)-\frac{1}{3a^2}\triangle\right]\left(\beta-\frac{1}{2}a^2\del_th\right), 
\label{eq:Kap}
\\
\Xi&:=&A-\frac{1}{a^2}\left(\beta-\frac{1}{2}a^2\partial_t h\right). 
\end{eqnarray}
The equations of motion can be reduced to the following equations:
\begin{eqnarray}
0&=&a\del_t\left(a\del_t\Phi\right)+3(1+w)a^2H\del_t\Phi 
-w \triangle\Phi 
+(1+w)a^2\Lambda \Phi, 
\label{eq:forPhi}\\
\Psi_{\rm lin}&=&-\Phi, 
\label{eq:PsiPhi}\\
\Delta&=&\frac{2}{a^2(3H^2-\Lambda)}\triangle \Phi, 
\label{eq:DeltaPhi}\\
\mathcal K&=&-\frac{1}{aH}\del_t(a\Phi), \\
V&=&-\frac{2}{(1+w)a^2(3H^2-\Lambda)}\del_t(a\Phi), 
\label{eq:VPhi}\\
\Xi&=&0. 
\end{eqnarray}
Therefore, a linear solution is expressed in terms of a solution of 
Eq.~\eqref{eq:forPhi}. 
For $w=\Lambda=0$, we obtain 
\begin{equation}
    0=\del_t(a^4\del_t\Phi)
    \label{eq:growconst}
\end{equation}
and find that the growing modes satisfy $\del_t\Phi=0$.

Another useful set of gauge-invariant variables is given by replacing $\mathcal K$ and $\Xi$ by 
$\widetilde{\mathcal K}$ and $\widetilde\Xi$ defined as 
\begin{eqnarray}
\widetilde{\mathcal K}&:=&\kappa-\left[\frac{1}{2}(1+w)\left(3H-\frac{\Lambda}{H}\right)-\frac{1}{3a^2H}\triangle\right]\left(\beta+v\right), \\
\widetilde \Xi&:=&A-\frac{1}{a^2}\left(\beta+v\right). 
\end{eqnarray}
By definition, these variables are equivalent to the perturbation of the trace and 
the traceless part of the extrinsic curvature in the comoving gauge $v+\beta=0$ 
or equivalently $u^\mu=n^\mu$. 
Therefore, these two variables correspond to the expansion and shear of the fluid four-velocity. 
From the equations of motion, we obtain 
\begin{eqnarray}
\widetilde{\mathcal K}&=&-\frac{2}{3}\frac{\del_t(a\triangle\Phi)}{(1+w)a^3H\left(3H^2-\Lambda\right)}, \\
\widetilde \Xi&=&2\frac{\del_t(a\Phi)}{(1+w)a^3\left(3H^2-\Lambda\right)}. 
\end{eqnarray}

Yet another useful variable is $\widetilde\Psi$ defined as follows:
\begin{equation}
\widetilde \Psi:=2\xi+H(\beta+v)-\frac{1}{6}\triangle h. 
\end{equation}
In the comoving gauge, we obtain 
% \AEE{I think a laplacian is missing in the next eqaution, it should be $\nabla^2 \widetilde \Psi $}\CY{right}
\begin{equation}
\triangle\widetilde \Psi=-\frac{1}{4}a^2\tilde R, 
\end{equation}
where $\tilde R$ is the Ricci scalar with respect to the 3-metric $\gamma_{ij}$. 
It can be shown that $\del_t\widetilde \Psi=0$ for $w=0$.

\section{Consistency between the BSSN simulation of collisionless particles
and the framework of the exact LTB solution
%Comparison between the simulation of the collisionless particle system and LTB
}
\label{sec:paandltb}

Here, we would like to show the consistency of the BSSN numerical simulation of a spherically symmetric system of collisionless particles free from velocity dispersion with the exact LTB solution of dust. This is not so easy as it looks at first sight because the gauge condition, such as slicing and threading,  adopted in the BSSN numerical simulation is highly nontrivial in terms of the coordinate system adopted for the description of the LTB solution. So, here, we will 
show that the spatial geometry on a time slice in the BSSN formulation is approximately the same 
as that properly calculated on the same spacelike hypersurface using the LTB exact solution.

To explicitly compare the geometry constructed by performing a BSSN numerical simulation and the corresponding LTB solution, let us consider a spacelike hypersurface $\Sigma$ 
in the coordinates $(t,r)$ in the LTB solution~\eqref{eq:LTBmetric}, 
given by 
\begin{equation}
\label{eq:inisli} 
%t=\dred{t_{\rm c}}+f(r), 
t=f(r),
\end{equation}
where the choice of the function $f(r)$ determines the hypersurface $\Sigma$.
We identify this hypersurface $\Sigma$ with a slice of time foliation in the BSSN numerical simulation.
Imposing that the time slice is smoothly connected to the standard cosmological time slice in the asymptotic region, 
we obtain the boundary condition $f'(r_{\rm b})=0$, with $r_{\rm b}$ being the radius of the outer boundary of the numerical domain.  

The unit normal form $n_\mu$ to this time slice is given by 
\begin{equation}
n_\mu=\Gamma\left(-(\dd t)_\mu+f'(\dd r)_\mu\right), 
\end{equation}
where
\begin{equation}
\Gamma=-n_\mu u^\mu=\left(1-{f'}^2\frac{1-kr^2}{(\del_r R)^2}\right)^{-1/2}. 
\label{eq:barE}
\end{equation}
From the definition of the extrinsic curvature $\nabla_\mu n^\mu=\frac{1}{\sqrt{-g}}\del_\mu\left(\sqrt{-g}n^\mu\right)=-K$, 
we obtain 
\begin{equation}
\frac{\sqrt{1-kr^2}}{\del_r R}
\frac{\dd}{\dd r}\left[\frac{\sqrt{1-kr^2}}{\del_r R}f'\Gamma\right]
+\left(\frac{2\del_t R}{R}+\frac{\del_t\del_rR}{\del_rR}
+\frac{2(1-kr^2)}{R\del_rR}f'\right)\Gamma=-K, 
\label{eq:forf}
\end{equation}
where 
\begin{equation}
\frac{\dd}{\dd r}:=\del_r+f'\del_t
\end{equation}
is the derivative with respect to $r$ on the hypersurface $\Sigma$. 
For practice, it is convenient 
to define new variable $A$ as follows:
\begin{equation}
A:=\frac{\sqrt{1-kr^2}}{\del_r R}f'\Gamma
=\left(\frac{(\del_r R)^2}{1-kr^2}{f'}^{-2}-1\right)^{-1/2}. 
\end{equation}
Then, Eq.~\eqref{eq:forf} can be decomposed into 
two first-order differential equations as follows:
\begin{eqnarray}
\frac{\dd}{\dd r} f &=&\frac{\del_rR A}{\sqrt{(1-kr^2)\left(A^2+1\right)}}, 
\label{eq:fprime}\\
\frac{\dd}{\dd r}A&=&-\frac{\del_rR}{\sqrt{1-kr^2}}\left[K+\left(\frac{2\del_tR}{R}
+\frac{\del_t\del_rR}{\del_rR}
+\frac{2(1-kr^2)}{R\del_rR}f'\right)\Gamma\right]. 
\label{eq:Aprime}
\end{eqnarray}
% \begin{eqnarray}
% f'&=&\frac{\del_rR A}{\sqrt{(1-kr^2)\left(A^2+1\right)}}, 
% \label{eq:fprime}\\
% A'&=&-\frac{\del_rR}{\sqrt{1-kr^2}}\left[K+\left(\frac{2\del_tR}{R}
% +\frac{\del_t\del_rR}{\del_rR}
% +\frac{2(1-kr^2)}{R\del_rR}f'\right)\bar \Gamma\right]. 
% \label{eq:Aprime}
% \end{eqnarray}
% Since the function $m$ is already fixed, all expressions in the right-hand side 
% can be calculated straightforwardly. 
% 
Once the functional form of $k(r)$, $m(r)$, and $K(r)$ is given, we can numerically solve Eq.~\eqref{eq:forf} under the condition 
$f'(0)=f'(r_{\rm b})=0$, 
where $R(t,r)$ and its derivatives 
on the right hand side of Eqs.~\eqref{eq:fprime} and \eqref{eq:Aprime}
are analytically evaluated using Eq.~\eqref{eq:ltbR} and \eqref{eq:defS} with
identification $t=f(r)$. 
When we solve it by a shooting method, we may consider $f(0)$ as the shooting parameter, and search for the value of $f(0)$ satisfying $f'(r_{\rm b})=0$. 
The functional forms of $k(r)$ and $m(r)$ are fixed by setting the functional form of $\Psi(r)$ through Eqs.~\eqref{eq:m} and \eqref{eq:k}. 

Performing the BSSN numerical simulations, we obtain a sequence of time slices, each of which can be characterized by the trace of the extrinsic curvature $K$. 
However, in the numerical simulation, the radial coordinate is not identical to the specific coordinate $r$. 
So, we use the proper length $l$ instead of the radial coordinate. We extract the value of $K$ as a function of the proper length $l$ from the origin on each time slice. 
On the LTB side, on the other hand, since the induced metric on the hypersurface $\Sigma$ is given by 
\begin{equation}
\dd l^2=\frac{\left(\del_r R\right)^2}{(1-kr^2)\Gamma^2}\dd r^2+R^2\dd \Omega^2, 
\end{equation}
the proper length can be calculated by integrating the following equation 
\begin{equation}
  \frac{\dd}{\dd r}l=\frac{\del_r R}{\sqrt{1-kr^2}\Gamma}. 
  \label{eq:lprime}
\end{equation}
Then we can perform numerical integration of Eqs.~\eqref{eq:fprime}, \eqref{eq:Aprime} and \eqref{eq:lprime}. So, even on the LTB side, we need the numerical integration to show the consistency of the BSSN result with the LTB description.

One of the simplest ways to compare the geometries of the time slices in the LTB solution and the numerical simulation
is to check the areal radius $R$ as a function of the proper length $l$ because both are coordinate-independent geometrical variables. 

% \subsection{Comparison}
We set $\mu=1.3$. 
Setting initial conditions generated from the functional form of $\Psi$, we perform the numerical simulation. 
Then we can obtain the areal radius $R$ as a function of the proper radius $l$ at each time step. 
The value of $K$ can be also extracted as a function of $l$. 
Substituting this functional form of $K$ into \eqref{eq:Aprime}, we can integrate Eqs.~\eqref{eq:fprime}, \eqref{eq:Aprime} and \eqref{eq:lprime}. 
Then we obtain the areal radius $R$ as a function of $l$ based on the LTB solution. 

\begin{figure}[h!]
\begin{center}
\includegraphics[scale=1.]{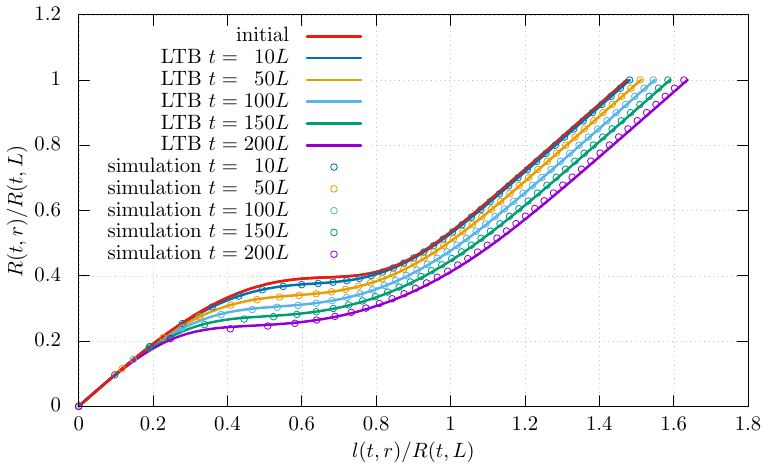}
\caption{    \baselineskip5mm
Functional forms of the areal radius $R$ as functions of the proper length $l$ on each time slice for 
the numerical simulation and the LTB solution with $\mu=1.3$. 
The number of grids for each side is set to 40. 
The solid lines show the results based on the LTB solution, and the circles show the results based on the numerical simulation. 
The horizontal and vertical axes are normalized by the areal radius for the numerical boundary given by $r=L$. 
}
\label{fig:lR}
\end{center}
\end{figure}

\begin{figure}[h!]
\begin{center}
\includegraphics[scale=1.]{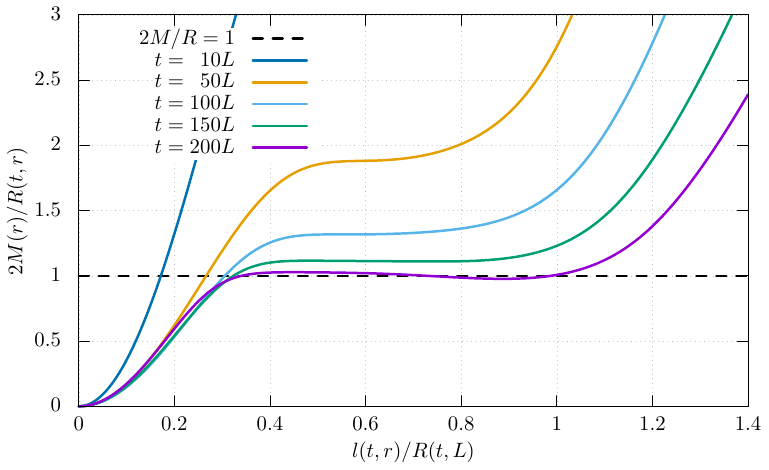}
\caption{    \baselineskip5mm 
The compactness defined by $2M/R$ as a function of the proper length $l$ evaluated on each time slice. 
}
\label{fig:lC}
\end{center}
\end{figure}

In Fig.~\ref{fig:lR}, we compare the functional forms obtained from the numerical simulation and the LTB solution, and 
one can see good agreement. 
We note that we introduce only 40 grid points for each direction in this demonstration. 
Even for this low resolution, we can find good agreements, which clearly demonstrate the robustness of the simulation 
as long as the fluid description remains valid. 
%%%%%%%%%%%%%%%%%%%%%%%%%%%<<start figure>>%%%%%%%%%%%%%%%%%%%%%%%%%%
%%%%%%%%%%%%%%%%%%%%%%%%%%%%<<end figure>>%%%%%%%%%%%%%%%%%%%%%%%%%%%
At the time $t=200L$, we could not find an apparent horizon due to the low resolution. 
However, as is shown in Fig.~\ref{fig:lC}, the line of the compactness $2M/R$ as a function of the proper length $l$ intersects 
the horizontal line of $2M/R=1$ three times. 
The outermost intersection point can be regarded as the cosmological horizon, and the middle one is the marginally outer trapped surface, which indicates the formation of a black hole. 
%%%%%%%%%%%%%%%%%%%%%%%%%%%<<start figure>>%%%%%%%%%%%%%%%%%%%%%%%%%%
%%%%%%%%%%%%%%%%%%%%%%%%%%%%<<end figure>>%%%%%%%%%%%%%%%%%%%%%%%%%%%

% \bibliographystyle{h-physrev5-title}
% \bibliography{MDPBH}

\end{document}